\documentclass[prb,superscriptaddress]{revtex4}
\usepackage{amsmath}                   % amsmath package
\usepackage{amsfonts}                  % amsfonts package
\usepackage{epsfig}
\usepackage{graphicx}                 

\newcommand{\lyxdot}{.}
\begin{document}
\author{Alexey A. Polotsky}
\affiliation{Institute of Macromolecular Compounds, Russian Academy of Sciences.
31 Bolshoy pr., 199004 St.-Petersburg, Russia}
\author{Tatiana M. Birshtein}
\affiliation{Institute of Macromolecular Compounds, Russian Academy of Sciences.
31 Bolshoy pr., 199004 St.-Petersburg, Russia}
\author{Mohamed Daoud}
\affiliation{Service de Physique de l'Etat Condens\'e CEA Saclay, 91191 Gif-sur-Yvette
Cedex, France}
\author{Oleg V. Borisov}
\affiliation{Institut Pluridisciplinaire de Recherche sur l'Environnement et les
Mat\'eriaux, UMR 5254 CNRS/UPPA, Pau, France}
\affiliation{Institute of Macromolecular Compounds, Russian Academy of Sciences.
31 Bolshoy pr., 199004 St.-Petersburg, Russia}
\email{oleg.borisov@univ-pau.fr}
\title{Conformations of amphiphilic polyelectrolyte stars with diblock copolymer arms}
%{Amphiphilic star diblock copolymers AB core shell star copolymer}

\begin{abstract}
We consider conformations and intra-molecular conformational transitions in 
amphiphilic starlike polymers formed by diblock copolymer arms with inner hydrophobic and outer polyelectrolyte blocks.
%in dilute aqueous solutions are considered. 
A combination of an analytical mean-field theory with the assumption-free numerical 
self-consistent field (SCF) modeling approach is applied. It is demonstrated that unimolecular micelles with collapsed hydrophobic cores and 
swollen polyelectrolyte coronae are formed in dilute aqueous solutions at high ionic strength or/and low degree of ionization of the outer hydrophilic block.
An intra-molecular conformational transition related to the unfolding of the hydrophobic core of the unimolecular micelles can be triggered by a decrease 
in the ionic strength of the solution or/and increase in the degree of ionization of the coronal blocks. In the stars with 
large number of diblock copolymer arms
the transition between conformations with
collapsed or stretched core-forming blocks occurs continuously by progressive unfolding of the core domain.
%and co-existence of collapsed and stretched 
%segments in each hydrophobic blocks. 
By contrast, in the stars with relatively small number of arms the continuous unfolding of the core
is interrupted by an abrupt unravelling transition. 
A detailed SCF analysis indicates that under both unfolding scenario 
the arms of the star are extended fairly equally, i.e., no intra-molecular disproportionation occurs.
\end{abstract}
\maketitle
\section{Introduction}
\begin{figure}[t]
\includegraphics[width=10cm]{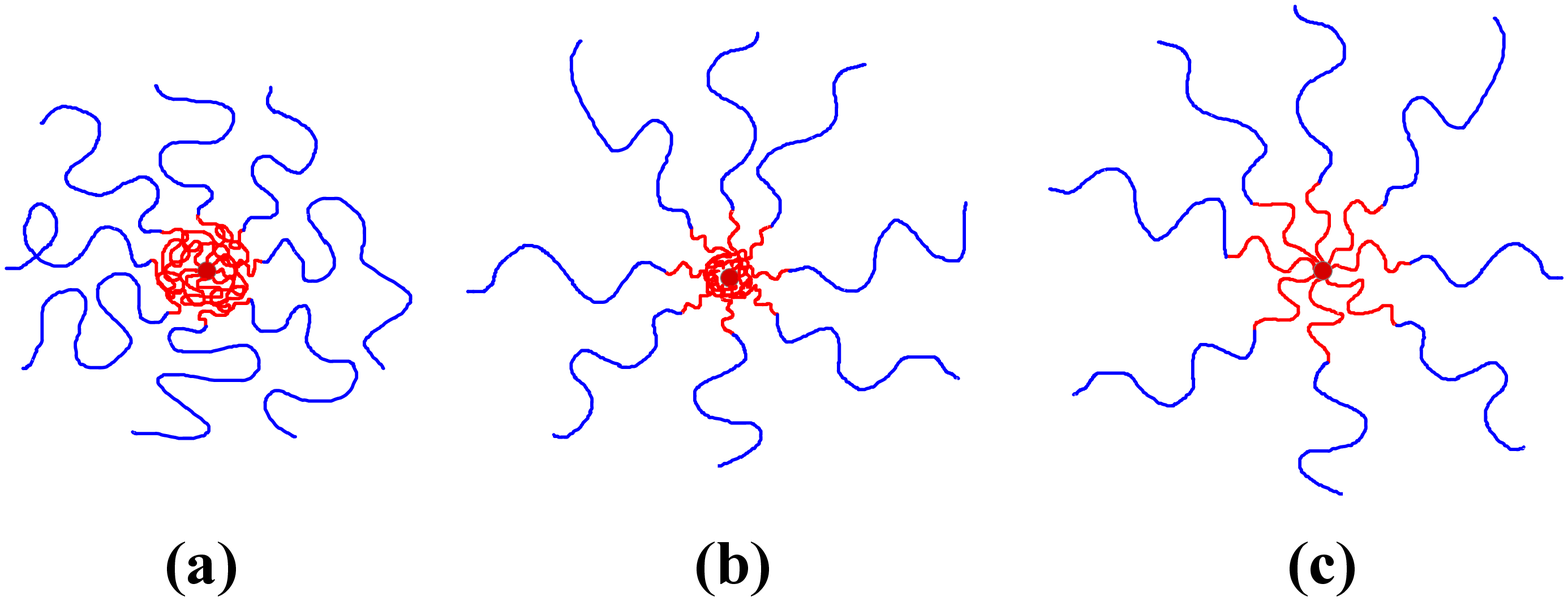}
\caption{\label{fig:AB_star}(AB)$_{p}$ core-shell star copolymer with collapsed (a), partially unfolded (b), and completely unfolded
(c) core}
\end{figure}

Micelles of amphiphilic block copolymers attract a lot of attention
in recent years because of their potential use in applications including
emulsion stabilization, viscosity regulation, biomedical applications,
and nanotechnology \cite{Riess:2003}.
One of the most interesting and promising fields of application of
amphiphilic block copolymers is their use for controlled drug delivery.
Many effective drug candidates have limited solubility, stability
and can be even rather dangerous because of their toxicity. To avoid
these transport problems, drug delivery systems on the basis of amphiphilic
block copolymer micelles with hydrophobic core and water-soluble corona
have been proposed \cite{Kataoka:2001, Gilles:2004}.
%%%%%%%%%%%%%%%%%%%%
%\bibitem{Kataoka} Kataoka K, Harada A, Nagasaki Y (2001) Advanced Drug
%Delivery Reviews 47: 113
%\bibitem{Gillies} Gillies ER, Fr\'echet MJ (2004) Pure Appl Chem 76: 1295
%%%%%%%%%%%%%%%%%%%%
In this case the core encapsulates a drug whereas
the hydrophilic corona makes this system soluble, protects from aggregation
and even makes it selective with respect to the pore
size thus allowing it to penetrate only into tumor tissues~\cite{Haag:2004}.
By attaching a specific terminal group to the corona chain, biocompatibility
and/or selectivity can be further enhanced. 
Upon reaching the target compartment, a proper release of the drug
molecules from the micelle should follow in response to the change
in the surrounding conditions. The latter in most cases is the pH
shift. For example, if the cellular uptake occurs via endocytosis~\cite{Alberts:2002,Haag:2004},
the pH changes from its physiological value pH 7.4 to sufficiently
lower value pH 4--5 in lysosome. Drug release can occur via dissociation
of the block copolymer micelle, cleavage of the corona chains or by
conformational changes in the core itself.
The use of starlike (AB)$_{p}$ block copolymers with hydrophobic inner B blocks and hydrophilic outer A blocks
which form in solution
the so called unimolecular micelles, Figure \ref{fig:AB_star}~a, has definite advantages over the
self-assembled AB diblock copolymer micelles. Unimolecular micelles
are more stable with respect to shear force, temperature, and pressure
(often used for sterilization) than the block copolymers aggregate
and, therefore, are more suitable for the noncovalent encapsulation
of guest molecules. Moreover, formation of the 
unimolecular micelles does not imply cooperative inter-molecular association, that is,
there is no CMC. 
Therefore, vector systems based on the unimolecular micelles can be used at arbitrary 
low concentration. 
Another advantage is the prescribed "aggregation
number", i.e. the number of star arms $p$ and thus, a better control of
the micelle size.This property is also very important for application of
(AB)$_{p}$ unimolecular micelles as nanoreactors for synthesis of
metal nanoparticles.
% and virtual independence of the critical micellisation
% concentration (CMC). 
%
At present, methods of synthesis of (AB)$_{p}$ stars are quite well
developed \cite{Fustin:2005} and stars with different chemical nature
of arm blocks solvable in water and organic solvents were obtained.
They were successfully used for synthesis of gold~\cite{Filali:2005}
and platinum~\cite{Zhang:2005} nanoparticles. First steps in the
creation of drug delivery systems on the basis of (AB)$_{p}$ unimolecular
micelles were also made~\cite{Haag:2004}.
The final and, therefore, crucial stage in the successful work of
a drug delivery system is the drug release when it reaches the target
compartment. Drug release can occur, as it was mentioned above, via
degradation of the core or of the micelle as a whole, but it may be
useful to keep the (AB)$_{p}$ star intact (for example, if the core
polymer itself cannot be safely removed from the body). This means
that the drug release should be made through unfolding of the globular
core, Figure \ref{fig:AB_star}~b, c. The driving force of this unfolding transition may be an increase
of the repulsive interactions in the corona (or a decrease in attractive
interactions in the core or both together) induced by variation of the environmental conditions. 
%Indeed, in the selective
%solvent, the core-forming block, for which the solvent is poor, tends
%to collapse and avoid the contacts with the solvent, whereas from
%the corona side, there is a force trying to extend the globule and
%thus win more space for swelling.
The corona of the (AB)$_{p}$ unimolecular micelle can be modeled as a
convex spherical polymer brush grafted onto the B-core and immersed into a good solvent. 
This system was the subject of intensive theoretical studies in the last
two decades beginning at the classical works of Daoud and Cotton~\cite{Daoud:1982},
Zhulina~\cite{Zhulina:1984:starconf,Zhulina:1984:stardiagr}, and
Birshtein and Zhulina~\cite{Birshtein:1984}. 
In more
recent work of Zhulina, Birshtein, and Borisov~\cite{Zhulina:2006} the Daoud-Cotton model
have been revised, generalized, and extended 
on the convex polyelectrolyte brush. 
%In these works, the
%expression of the brush free energy as the function of the core radios
%and grafting density have been obtained, from what the force dependence
%can be calculated straightforwardly. 
It should be noted that these
approaches employed the so called "fixed ends" (or equal stretching) approximation, 
consisting in the assumption that the free ends of the grafted chains
are localized at the edge of the brush, hence, all chains are equally,
although not uniformly, stretched. For planar polymer brushes this approximation is
often referred to as Alexander--de Gennes approximation~\cite{Alexander:1977, DeGennes:1980}.

The collapsed core of the unimolecular micelle can be also considered
as a spherical polymer brush grafted to a small sphere with the free
ends fixed at the outer boundary of the brush (hence, the core fits
well to the Alexander-de Gennes approximation). Coil-globule transition
in the star-shaped polymer in poor solvent was studied by Zhulina
\emph{et~al.}~\cite{Zhulina:1988} using scaling approach. 
%and Alexander-de Gennes approximation. 
It was demonstrated that under poor solvent conditions the intra-molecular concentration
profile consists of two parts: 
%The inner $\theta$-region is dominated by ternary monomer-monomer repulsions,
a denser inner (central) zone is due to tethering of the star arms to the central point,
here the polymer concentration decays inversely proportional to the distance from the star center.
In the rest of the star the concentration is constant and is equal
to the characteristic concentration of the collapsed polymer globule.
%($\sim|\tau|$,
%where $\tau$ is the relative deviation from $\theta$-temperature,
%$\tau\equiv(T-\theta)/T$) \cite{deGennes:1979}.

The core-forming B-blocks of the amphiphilic star copolymer are subjected to the radial pulling force
exerted by crowded coronal blocks. When this force exceeds a certain critical threshold,
it may induce unraveling of the collapsed core and stretching of the B-blocks in the radial direction.
%As follows from simple scaling arguments presented below, in the case of non-ionic coronal blocks 
%swollen in a good solvent the pulling force is not strong enough to induce unraveling of the collapsed core.
In particular, the long-range Coulomb interactions between ionic (polyelectrolyte) A-blocks under 
low ionic strength conditions may be strong enough to provoke such kind of conformational transition.
In~\cite{Mercurieva:2009}, conformational characteristics of amplhiphilic (AB)$_5$ star 
(with block lengths $N_A=80$, $N_B=20$ and fraction of charged units in the A block $\alpha_b=0.5$) 
as functions of ionic strength were studied by means of numerical self-consistent field modeling. 
It was shown that at high salt concentrations 
%($\varphi_s=10^{-1}$) 
all 
B-blocks are not extended by charged A-blocks and collapse in the core while at low salt concentrations 
%($\varphi_s=10^{-4}$) 
the B-blocks are strongly stretched due to the corona pulling force.
Monte Carlo simulation of (AB)$_{p}$ stars were performed by Nelson
\emph{et. al.}~\cite{Nelson:1998} who considered not only the star
with solvophobic core and solvophilic corona but the opposite interesting
case (solvophilic inner + solvophobic outer blocks), paying more attention
to the latter one. In these simulations, 
%solvophilic monomers were identical (from the point of view of interactions) to the solvent molecules (i.e. 
solvent was athermal for the solvophilic block, 
%Flory parameter $\chi=0$ 
and the Flory parameter for the outer block
was varied. It was found that the stars 
which have 
%having the solvophilic end of the di
the innter solvophilic block can undergo a dramatic
conformational change in which the outer solvophobic blocks aggregate
into one or more compact globules. This aggregation transition
is accompanied by a significant change in the size of the polymer
as measured by the radius of gyration. In the case of 
stars in which the inner blocks are solvophobic and the outer blocks
are solvophilic like in our study, no such transition was found.

The aim of the present work, is to study equilibrium conformations of the 
amphiphilic (AB)$_{p}$ star block copolymers with hydrophobic B-blocks and strongly dissociating
(quenched) polyelectrolyte A-blocks. We are particularly interested in the analysis of 
the stimuli-induced conformational
(unfolding) transition in the (AB)$_{p}$ star triggered by the change in
the ionic strength of the solution or/and degree of ionization of the A-blocks,
which affect the strength of repulsive interactions in the corona.
%the effective interaction in the A-corona and B-core. 
%As an important
%part of this problem, theory of deformation of a collapsed (globular)
%star will be developed. The latter will take into account the microsegregation
%in the deformed star into central globular part and stretched {}``legs''
%similarly to that in the deformed polymer globule.

The rest of the paper is organized as follows: in Section ``Model and general considerations'' we specify our model of the 
(AB)$_{p}$ star block copolymer. In Section ``Polyelectrolyte coronal domain'' we calculate the pulling force exerted by the corona A-blocks onto the core blocks as a function
of variable charge parameters and ionic strength of the solution. 
In Section ``Core domain'' we consider the B-core domain and analyze 
conformation of a collapsed in poor solvent star homopolymer subjected to radial extensional deformation.
Our main results in terms of the diagrams of states and
analysis of the character of the intra-molecular unfolding transition in the (AB)$_{p}$ star block copolymer
are presented in Section ``Conformational transition in (AB)$_{p}$ star block copolymer''. 
The analytical results are compared to the detailed self-consistent field (SCF) modeling in Section ``SCF modeling''.
In the end, we summarize our conclusions and present an outlook for the future work.
\section{Model and general considerations}
We consider a dilute aqueous solution of (AB)$_{p}$ star copolymers.
A star comprises $p$ AB diblock copolymer arms, each consisting of
a hydrophobic block B with the degree of polymerization $N_{B}$ and
a polyelectrolyte block A with the degree of polymerization $N_{A}$.
The arms are
grafted by the ends of the B-blocks to a small sphere with the radius
$r_{0}$. Both blocks are assumed to be intrinsically flexible. That
is, the statistical segment length is of the order of a monomer unit
length $a$.
The block A comprises a fraction $\alpha_b$ of permanently 
charged monomer units, i.e. is a quenched polyelectrolyte.
Water is assumed to be a marginal good solvent for both charged and uncharged monomer units of the
A-block. The short-range
interactions between A monomers are modeled in terms of the virial
expansion and only the pair monomer-monomer interactions with second
virial coefficients $v_{A}$ are considered to be relevant. 
On the contrary, for the hydrophobic block B water is a poor solvent,
the short-range (van der Waals) interactions between B-monomers are
modeled in terms of Flory-Huggins parameter $\chi_{B}$,
under poor solvent conditions $\chi_{B}>0.5$.
Below we consider only very dilute solution of (AB)$_{p}$ star copolymer
where 
%the molecules do not aggregate and form unimolecular micelles only. 
the interactions between the stars are weak and do not affect
their conformations. 
%In the case when the solution is not dilute, aggregation
%of (AB)$_{p}$ star copolymers may lead to formation of spherical
%micelles with multiple ($2p$, $3p$, $4p$, ...) number of arms or
%to morphology change (cylindrical micelles, lamellar or bicontinuous
%structure) but this situation lies beyond the scope of the present
%investigation.
To develop an analytical theory we use a simple model assuming that all hydrophobic B-blocks are equally (but not uniformly) stretched so all the A-B junction points are located at the
interface on the equal distance $R$ from the center of the star. Then all hydrophobic B-blocks are within the sphere of the radius $R$ while the coronal A-blocks form a swollen brush on the surface of this sphere. This approximation is verified and justified below by detailed assumption-free numerical SCF calculations.
Equilibrium value of $R$ corresponds to the minimum of the free energy of the star:
\begin{equation}
\label{free_energy}
F(R)=F_A(R)+F_B(R),
\end{equation}
where $F_X$ are contributions of hydrophilic ($X=A$) and hydrophobic ($X=B$) blocks. Minimization of the free energy, eq \ref{free_energy}, is equivalent to the force balance condition
\begin{equation}
\label{force_balance}
\mathfrak{f}_{A}(R)+\mathfrak{f}_{B}(R)=0
\end{equation}
where $\mathfrak{f}_{X}(R)=dF_X(R)/dR$ are the forces acting from the block $X$ ($X=A,\, B$) on the phantom surface separating the blocks.
In the absence of the corona hydrophobic B blocks collapse and form spherical core with the radius $R_0$. This core preserves its shape in the amphiphilic star if the corona pulling force $\mathfrak{f}_A(R_0)$ is not sufficiently strong. With increasing pulling force (for example, caused by a decrease in the ionic strength) it reaches a certain threshold value which is enough to surmount the core restoring force corresponding at least to partial unfolding of the hydrophobic core. Obviously, this threshold is higher for more hydrophobic core. As it will be demonstrated below, in a wide range of pulling forces, inner hydrophobic part of the (AB)$_p$ star has the shape of an ``octopus'' (compare with~\cite{Williams:1993}) with reduced dense spherical core and stretched B branches similar to extended hydrophilic corona branches. At very high pulling forces globular core disappears and the B-blocks acquire stretched in the radial direction conformation. 

In the following two sections we consider separately the swollen polyelectrolyte corona and hydrophobic core domain and
derive the corresponding free energy as well as pulling and reaction force-radius dependences. The obtained dependences will be then used for analysis of equilibrium conformation of the (AB)$_p$ star as a whole.
\section{Polyelectrolyte coronal domain}
\label{sec:corona}To calculate the corona free energy, we use the
approach developed by Zhulina~\emph{et al.}~\cite{Zhulina:2006}.
Consider a polymer brush composed by $p$ chains uniformly grafted
onto impenetrable sphere with the radius $R$.
The free energy of the brush (per chain) can be expressed as \begin{equation}
\frac{F}{k_{B}T}=\frac{3}{2a^{2}}\int_{R}^{D}\left(\frac{dr}{dn}\right)dr+\int_{R}^{D}f_{int}\left\{ c_{p}(r)\right\} s(r)\, dr,
\label{eq:F_gen}
\end{equation}
where $f_{int}$ is the free energy density of interactions in $k_B T$ units.
The latter equation implies that the end segments of all the A-blocks are located at distance $D$ from the centre of the star,
i.e. equal (but non-uniform !) stretching of the A-blocks is pre-assumed. 
The first term describes elastic stretching of the chain. The local
chain extension $dr/dn$ is related to the local polymer concentration
$c_{p}(r)$ \begin{equation}
c_{p}(r)=\frac{dn}{s(r)\cdot dr}=\frac{dn}{dr}\cdot\frac{p}{4\pi r^{2}}.\label{eq:c}\end{equation}
$s(r)$ is the area per chain at the distance $r$ from the center
of the grafting sphere: \begin{equation}
s(r)=\frac{4\pi}{p}\cdot r^{2}.\label{eq:sigma_star}\end{equation}
The second term in eq \ref{eq:F_gen} describes excluded volume and
electrostatic interactions within the star and can be represented
as \begin{equation}
f_{int}\left\{ c_{p}(r)\right\} =f_{ev}\left\{ c_{p}(r)\right\} +f_{ion}\left\{ c_{p}(r)\right\} \label{eq:fint_corona}\end{equation}
where \begin{equation}
f_{ev}\left\{ c_{p}(r)\right\} =v_{A}c_{p}^{2}(r)\label{eq:fev_corona}\end{equation}
accounts for the excluded volume interactions in the corona.
To specify the electrostatic contribution $f_{ion}\left\{ c_{p}(r)\right\} $
for ionic corona, the combination of the mean-field and the local
electroneutrality approximation (LEA) is used. 
The LEA assumes that the local charge of polyelectrolyte coronal chains is compensated by the
excess local concentration of counterions (including oppositely charged ``own'' polyelectrolyte counterions and salt ions). All ions in the system are monovalent. The LEA is applicable provided
that the number of block copolymer arms in one star is sufficiently large so
that the excess electrostatic potential is great enough to retain the mobile
counterions inside the corona, even at low salt concentrations in the
solution. In the LEA framework, the electrostatic interactions manifest
themselves through the translational entropy of the ions, disproportionated between the
interior of the corona and the bulk solution. The concentrations (the chemical potentials)
of all mobile ions in the bulk of the solution are assumed to be fixed.
The distribution of mobile co- and
counterions between interior of the corona and the bulk solution is
determined according to Donnan rule. The details of the approach can
be found in~\cite{Zhulina:2006}. For strong (quenched)
polyelectrolyte with constant degree of ionization $\alpha_{b}$ it
is given by~\cite{Zhulina:2006} \begin{equation}
\begin{split}f_{ion}\left\{ c_{p}(r)\right\} = & -\left(\sqrt{1+(\alpha_{b}c_{p}(r)/\Phi_{ion})^{2}}-1\right)\Phi_{ion}\\
& -\alpha_{b}c_{p}(r)\ln\left(\sqrt{1+(\alpha_{b}c_{p}(r)/\Phi_{ion})^{2}}-\alpha_{b}c_{p}(r)/\Phi_{ion}\right).\end{split}
\label{eq:fion_corona}\end{equation}
where $\Phi_{ion}=\sum_{j}c_{bj}$ is the bulk concentration of all
monovalent ions species.
Therefore, the free energy can be expressed as \begin{equation}
\frac{F}{k_{B}T}=\int_{R}^{D}f\left\{ c_{p}(r),r\right\} s(r)\, dr\,,\label{eq:F_star}\end{equation}
where \begin{equation}
f\left\{ c_{p}(r),r\right\} =\frac{3}{2a^{2}}\cdot\left(\frac{p}{4\pi}\right)^{2}\cdot\frac{1}{r^{4}c_{p}(r)}+
f_{int}\left\{ c_{p}(r)\right\} \label{eq:fd_star}\end{equation}
is the free energy density.
The outermost radius of the corona $D$ is related to the polymer concentration profile
$c_{p}(r)$ via the constraint of conservation of the total number
of monomers \begin{equation}
\int_{R}^{D}c_{p}(r)\, s(r)\, dr=\frac{4\pi}{p}\int_{R}^{D}c_{p}(r)\, r^{2}\, dr=N_{A}\,.\label{eq:norm_gen}\end{equation}
In the following we employ the so called local or quasi-planar (QP)
approximation \cite{Zhulina:2006} where instead of minimization of
the functional, eq \ref{eq:F_star}, with respect to $c_{p}(r)$ the
local condition of vanishing the differential osmotic pressure in
the corona: 
\begin{equation}
\pi(r)=c_{p}^{2}(r)\frac{\delta}{\delta c_{p}(r)}\left(\frac{f\left\{ c_{p}(r),r\right\} }{c_{p}(r)}\right)=0\label{eq:osmpr}
\end{equation}
is used. This leads to the following equation for $c_p(r)$:
\begin{equation}
-\frac{3}{a^{2}}\left(\frac{p}{4\pi}\right)^{2}\frac{1}{r^{4}c_{p}^{3}(r)}+v_{A}+\Phi_{ion}\frac{\sqrt{1+(\alpha_{b}c_{p}(r)/\Phi_{ion})^{2}}-1}{c_{p}^{2}(r)}=0.\label{eq:localcond_corona}\end{equation}
from what the polymer density profile follows. Explicitly it can
be found only in the {}``inverse form'' $r=r(c_{p})$ \begin{equation}
r=\left(\frac{p}{4\pi}\right)^{1/2}\frac{3^{1/4}}{a^{1/2}c_{p}^{1/4}[v_{A}c_{p}^{2}+\Phi_{ion}(\sqrt{1+(\alpha_{b}c_{p}/\Phi_{ion})^{2}}-1)]^{1/4}}.\label{eq:rc_profile_corona}\end{equation}
Then, the normalization condition, eq \ref{eq:norm_gen}, can be re-written
as an integral over $c_{p}$:
\begin{equation}
\frac{4\pi}{p}\int_{R}^{D}c_{p}(r)\, r^{2}\, dr=
\frac{4\pi}{p}\int_{c_{p}(R)}^{c_{p}(D)}c_{p}r^{2}(c_{p})\,\frac{dr(c_{p})}{dc_{p}}\, dc_{p}=N_{A}\label{eq:norm_cp}\end{equation}
The free energy functional can be written as \begin{equation}
\frac{F_{A}}{k_{B}T}=\frac{4\pi}{p}\int_{R}^{D}f\left\{ c_{p}(r),r\right\} \, r^{2}\, dr=
\frac{4\pi}{p}\int_{c_{p}(R)}^{c_{p}(D)}f(c_{p})r^{2}(c_{p})\, \frac{dr(c_{p})}{dc_p}\, dc_{p}.\label{eq:F_cp}\end{equation}
where $f(c_{p})$ is obtained from $f\left\{ c_{p}(r),r\right\} $,
eq \ref{eq:fd_star}, by using eq \ref{eq:rc_profile_corona}
\begin{equation}
\begin{aligned}f(c_{p})= & \frac{3}{2}v_{A}c_{p}-\frac{1}{2}\Phi_{ion}\left[\sqrt{1+(\alpha_{b}c_{p}/\Phi_{ion})^{2}}-1)\right]\\
& -\alpha_{b}c_{p}\ln\left[\sqrt{1+(\alpha_{b}c_{p}/\Phi_{ion})^{2}}-\alpha_{b}c_{p}/\Phi_{ion})\right]\end{aligned}
\label{eq:fcp}\end{equation}
We can obtain the dependence of the free energy $F_{A}$ as a function
of the core radius $R$ as follows. Polymer concentration at the core
surface, $c_{p}(R)$, is related to the core radius $R$ via eq \ref{eq:localcond_corona}
or eq \ref{eq:rc_profile_corona}. Once parametrically $c_{p}(R)$ is
known, the polymer concentration at the periphery of the brush, $c_{p}(D)$,
is found using normalization condition, eq \ref{eq:norm_cp}, then, finally
we use eqs \ref{eq:F_cp} and \ref{eq:fcp} to find $F_{A}$.
%Within the local approximation the elastic tension in the coronal A block at distance $r$ from the center of the star
%is given by 
%\begin{equation}
%\mathfrak{f}_{A}(r)=k_{B}T\cdot\frac{3p}{4\pi r^2}\cdot\frac{1}{c_p(r)}=k_BT
%\biggl[\frac{3\Phi_{ion}}{a^2c_p(r)}\biggl(\sqrt{1+\left(\frac{\alpha c_p(r)}{\Phi_{ion}}\right)^2}-1\biggr)\biggr]^{1/2}
%\end{equation}
The stretching (pulling) force from the corona side applied to the B-blocks at the A-B junction points
is expressed as 
% within the local approximation
%\begin{equation}
%\mathfrak{f}_{A}(R)=k_{B}T\cdot\frac{3p}{4\pi R^2}\cdot\frac{1}{c_p(R)}
%\end{equation}
\begin{equation}
\begin{aligned}
\mathfrak{f}_{A} = \frac{dF_{A}}{dR} = k_{B}T\cdot\frac{4\pi}{p}\left[f\left\{ c_{p}(D)\right\} D^{2}\,
\frac{\partial D}{\partial R}-f\left\{ c_{p}(R)\right\} R^{2}\right] \\
= - k_{B}T\cdot\frac{4\pi}{p} R^2 f\left\{ c_{p}(R)\right\} \left[1- \frac{f\left\{ c_{p}(D)\right\}}{f\left\{ c_{p}(R)\right\}} \cdot \frac{c_{p}(R)}{c_{p}(D)}\right]
\end{aligned}
\end{equation}
$\partial D/\partial R$ is found by differentiating the normalization condition~eq~\ref{eq:norm_gen} with respect to $R$.
%\begin{equation}
%\frac{4\pi}{p}\int_{R}^{D}c_{p}(r)\, r^{2}\, dr=N_{A}\,.\label{eq:norm_corona}\end{equation}
%with respect to $R$: \begin{equation}
%c_{p}(D)D^{2}\,\frac{\partial D}{\partial R}-c_{p}(R)R^{2}=0\end{equation}
%from what follows 
%\begin{equation}
%\mathfrak{f}_{A}=k_{B}T\cdot\frac{4\pi}{p}\, c_{p}(R)R^{2}\left(\frac{f\left\{ c_{p}(D)\right\} }{c_{p}(D)}-\frac{f\left\{ c_{p}(R)\right\} }{c_{p}(R)}\right).
%\end{equation}
\begin{figure}[t]
(a) \includegraphics[width=7cm]{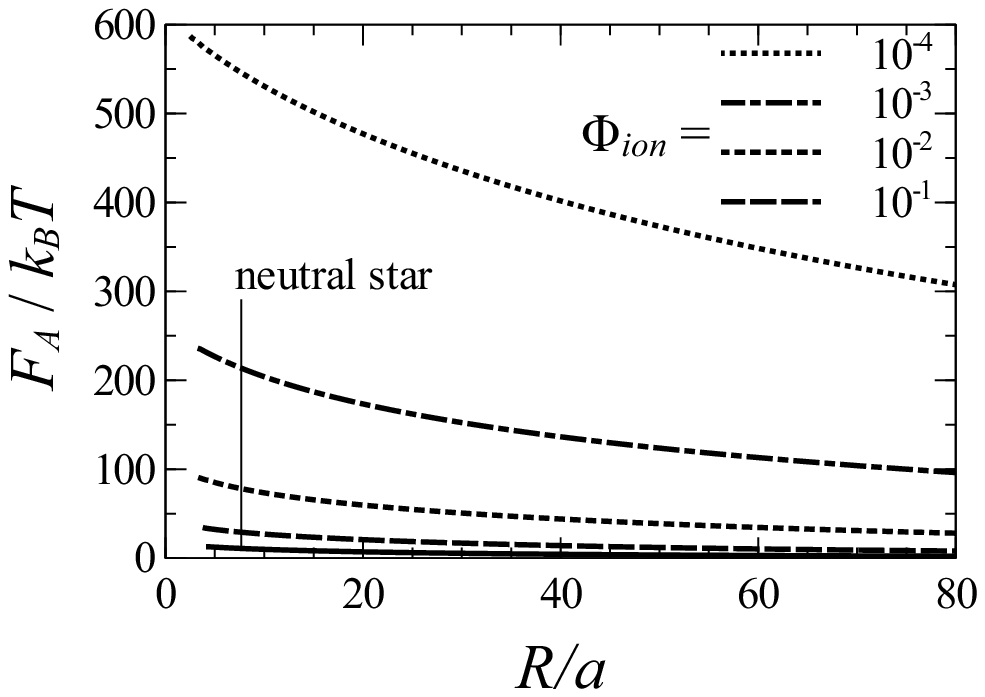}
\includegraphics[width=7cm]{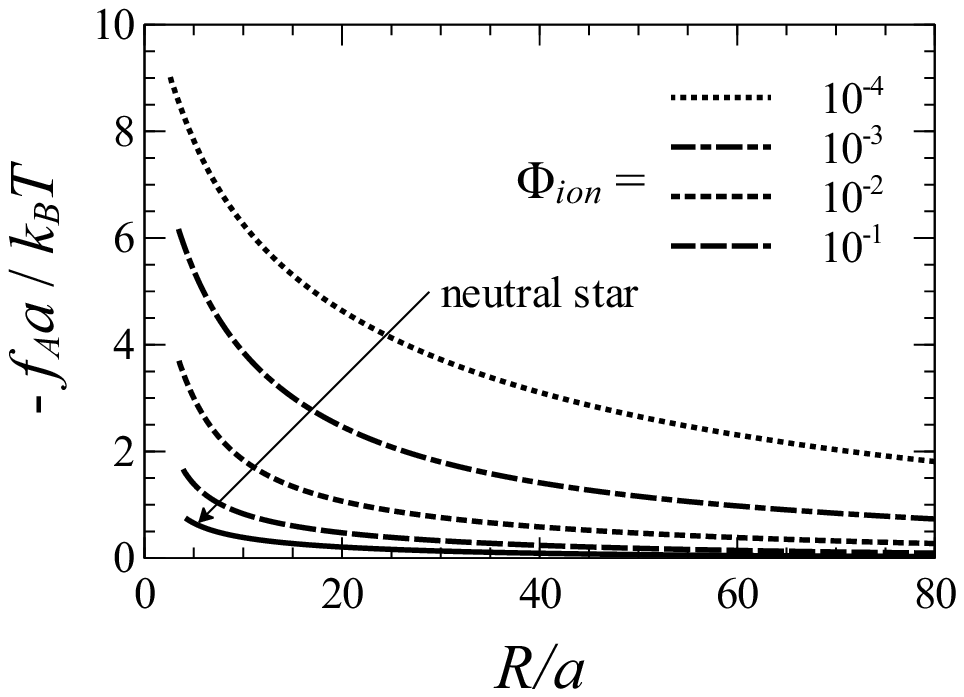} (b)
\caption{\label{fig:Fcorona}Free energy of polyelectrolyte corona
(per chain) (a) and corresponding pulling force (b) as functions of
the core radius for different salt concentrations, $\phi_{ion}\equiv \Phi_{ion}a^3$.
The values of parameters are $N_{A}=1000,$ $p=20$, $\alpha_{b}=0.5$,
$v_{A}=a^3/6$}
\end{figure}
Figure \ref{fig:Fcorona} shows the free energy and pulling force dependences
of the radius $R$ calculated for $N_{A}=1000,$ $p=20$, $\alpha_{b}=0.5$,
$v_{A}=a^3/6$ and various volume fractions of salt, $\phi_{ion}\equiv \Phi_{ion}a^3$. Both $F_{A}$ and $\mathfrak{f}_{A}$
are decreasing functions of $R$, a decrease in the ionic strength leads to an
increase of the pulling force $\mathfrak{f}_{A}$.
\section{Core domain}
The core domain of the (AB)$_{p}$ star can be considered
as a polymer star under poor solvent conditions with each arm
subjected to radial tension exerted by the swollen
coronal blocks. The analysis of the conformations of the core blocks necessitates taking into account the specific effects
appearing upon extensional deformation of a polymer globule. The latter
reveals, in its turn, an non-trivial physical picture.
\subsection{Extensional deformation of a single chain in poor solvent}
As it was first
shown by Halperin and Zhulina \cite{Halperin:1991:EL,Halperin:1991:MM}, when a constant force is applied to the ends
of a polymer chain collapsed in poor solvent, there exists a critical force, below which the globule
remains compact but changes its shape from spherical to prolate ellipsoidal,
whereas above the critical force it unwinds into a stretched chain.
When the globule is deformed in the extension ensemble,
i.e. the distance between chain ends is imposed, the unfolding scenario is more complex: it occurs via formation of the ``tadpole'' conformation in which the spherical
globular ``head'' coexists with the stretched string of thermal blobs (``leg'').
%at small deformations
%the globule changes its shape and the restoring force is proportional
%to the extension, $\mathfrak{f}\sim L$. At moderate extensions the macromolecule
%acquires the "tadpole" conformation in which the spherical
%globular "head" coexists with the stretched string of thermal blobs ("leg"). The
%restoring force in this "tadpole" regime is approximately constant
%independently of the distance between chain ends, $\mathfrak{f}\sim L^{0}$.
%(According to more accurate estimates the force even slightly decays
%as a function of stretching). The number of monomer units in the globule decreases
%and, correspondingly, the number of monomer units in the tail increases
%upon extension. Then, for the strong stretching, the linear response
%$\mathfrak{f}\sim L$ is recovered, up to finite extensibility limit. Later on
%this approach has been extended to the case of collapsed semirigid
%chain \cite{Craig:2005:1}. The predicted picture of globule deformation
%was also confirmed in simulations \cite{Frisch:2002,Cooke:2003,Cieplak:2004},
%although in the case of lattice models or when the formation of helical
%conformation is possible a multistep transition can be observed \cite{Marenduzzo:2003,Marenduzzo:2004}
In our recent works \cite{Polotsky:2009,Polotsky:2010} we have performed
a detailed self-consistent field (SCF) calculations and developed a
quantitative analytical theory of the equilibrium unfolding of a globule
formed by a flexible homopolymer chain collapsed in a poor solvent
and subjected to an extensional deformation. Both the SCF calculations
and the analytical theory prove that upon an increase
in the end-to-end distance three regimes of deformation successively
occur: at small deformation the globule acquires a prolate shape,
the reaction force grows linearly with the deformation; at moderate
deformations the globule is in the tadpole conformation, the tail of the tadpole is uniformly stretched, its extension as well as the reaction
force is weakly decreasing with the deformation; then at certain extension the globular
head unravels and the globule completely unfolds, the reaction force
drops down and then grows again upon further extension (this jumpwise ``unraveling transition'' was first predicted by Cooke and Williams~\cite{Cooke:2003}). In the framework
of the developed theory it was possible to calculate prolate globule-tadpole
and tadpole-open chain transition points and to find corresponding
conformational characteristics, such as the asymmetry and the number
of monomers in the globular head, reaction forces and force jump at
transition, in a wide range of polymerization degree $N$ and solvent
quality ($\chi$). Our analysis has shown that the system exhibits
a critical point; i.e., there exists a minimal chain length $N_{cr}(\chi)$
below which, at $N<N_{cr}(\chi)$, the intramolecular microphase segregation
in the extended globule does not occur. The globule is deformed 
as a whole, without intramolecular segregation but by progressive 
depletion
of its core.
\subsection{Homopolymer star in poor solvent}
\label{sec:core}Consider a free polymer star comprising $p$ arms each
of $N_{B}$ monomer units immersed into a poor solvent, $\chi\geq 0.5$. 
The star can be
represented as a spherical polymer brush grafted onto a small sphere with
the radius $r_{0}$. The mean grafting area per chain is then $s_{0}=4\pi r_{0}^{2}/p$.
Since $p$ arms cover the surface of the sphere and the arm cross-section
area equals to $a^{2}$, $r_{0}$ can be estimated as $r_{0}=a\sqrt{p/(4\pi)}$.
It is assumed that all the chains are extended equally but non-uniformly
in the radial direction so that their free ends are localized at the
edge of the star at the distance $R_{0}$ from its center.
%(Alexander-deGennes approximation~\cite{Alexander:1977,DeGennes:1980}). 
It can
be shown~\cite{Zhulina:1988} that the concentration profile of the
collapsed star consists of three concentric regions: Close to the center, 
there is a narrow zone where polymer volume fraction is approximately
equal to unity. Then
% in the so called $\theta$-zone 
it decays as $1/r$ down to the concentration $\varphi$ which is the characteristic
globular concentration. In the Flory-Huggins model, the polymer concentration
$\varphi$ or, equivalently, the volume fraction in the globule $\phi\equiv\varphi a^{3}$
can be expressed via Flory parameter $\chi$ as \cite{Polotsky:2010}
$\chi\simeq-\frac{\ln(1-\phi)}{\phi^{2}}-\frac{1}{\phi}$. 
Since
the radius of the central zone of decaying concentration is small compared to the radius of the globule
$R_{0}$, 
in the following calculation we assume that the polymer
concentration in the collapsed star is uniform and equal to $\varphi$, i.e.
\begin{equation}
\label{eq:R_0}
R_{0}\simeq \left(r_{0}^{3}+\frac{3N_{B}p}{4\pi\varphi}\right)^{1/3}
\end{equation}
This
simplification should not affect, neither qualitatively nor quantitatively,
final result.
\begin{figure}[t]
\includegraphics[width=10cm]{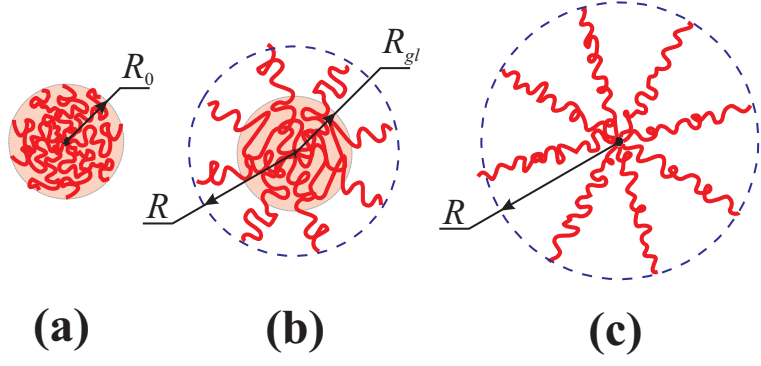}
\caption{\label{fig:B_Star} Polymer star collapsed in poor solvent: (a) free
(unperturbed) star, (b) radially stretched star at moderate deformation,
(c) strongly radially stretched star.}
% {\bf Shall we show in this figure $R_0$ and $R$}}
\end{figure}
Consider now the situation where the ends of star arms are "pinned"
at the distance $R\geq R_{0}$, i.e. each arm of the collapsed star is radially stretched. In
accordance with the general picture of polymer globule deformation
\cite{Halperin:1991:EL,Polotsky:2009} we propose the following model:
we assume that under such extensional deformation, the star segregates
into corona formed by equally stretched "tails" and spherical globular
core, Figure \ref{fig:B_Star}~b. The polymer concentration in the core
is still equal to $\varphi$ whereas the tails are the uniformly stretched
parts of the polymer chains exposed to the solvent. If the core contains
$N_{gl}$ monomer units per arm and has, therefore, the radius 
\begin{equation}
R_{gl}=\left(r_{0}^{3}+\frac{3N_{gl}p}{4\pi\varphi}\right)^{1/3},
\label{eq:Rgl}
\end{equation}
then the free energy of the deformed star \emph{per star arm} can
be represented as follows 
\begin{equation}
\frac{F_{B}}{k_{B}T}=\frac{3}{2a^{2}}\int_{r_{0}}^{R_{gl}}\left(\frac{dr}{dn}\right)dr+\mu N_{gl}+
\gamma\cdot\frac{4\pi R_{gl}^{2}}{p}+\frac{3(R-R_{gl})^{2}}{2(N_{B}-N_{gl})a^{2}}\label{eq:Fcore0}
\end{equation}
where the first term is the elastic free energy of the globular core ($r$
is the radial coordinate and $n$ is the monomer ranking number counted within each arm from the center
of the star),
$k_BT\mu$ is the monomer chemical potential (free energy) in the infinite
globule ($\mu$ is negative under poor solvent conditions), the third
term accounts for the interfacial contribution from the surface of
the globule, $k_BT\gamma$ is the interfacial tension coefficient ($\gamma$
is positive due to the energetic and conformational penalty of redundant
monomer-solvent contacts at the interface), and the last term is the
elastic free energy of the tail. In the volume approximation, which is applicable for sufficiently large core, it is
assumed that both $\mu$ and $\gamma$ are functions of the solvent
quality only (i.e. independent of the core size), their dependences
on the Flory parameter $\chi$ are expressed as follows: the monomer
chemical potential $\mu$ can be found, once the polymer volume fraction
within the globule, $\phi$, is known \cite{Polotsky:2010}:
\begin{equation}
\mu=2+\frac{2-\phi}{\phi}\log(1-\phi).\end{equation}
For the interfacial tension coefficient $k_BT\gamma$ an explicit expression
can be obtained for moderately poor solvent only \cite{Ushakova:2006, Polotsky:2010}:
\begin{equation}
\gamma a^{2}=\frac{3}{16}(1-2\chi)^{2}\end{equation}
Comparison with the values of $\gamma$ obtained numerically using
Scheutjens-Freer self-consistent field (SF-SCF) approach shown \cite{Polotsky:2010}
that this approximation works good for $\chi\lesssim1.2$ which is
the range of $\chi$ we are interesting in.
%because the
%core should not be too dense in order to be unfolded by the corona
%of the diblock copolymer star).
Since the polymer concentration in the globule is constant and equal
to $\varphi$, local chain tension is $dr/dn=p/(4\pi r^{2}\varphi)$,
and the integral in eq \ref{eq:Fcore0} can be easily calculated \begin{equation}
\frac{F_{B}}{k_{B}T}=\frac{3p}{8\pi a^{2}\varphi}\left(\frac{1}{r_{0}}-\frac{1}{R_{gl}}\right)+\mu N_{gl}+\gamma\cdot\frac{4\pi R_{gl}^{2}}{p}+\frac{3(R-R_{gl})^{2}}{2(N_{B}-N_{gl})a^{2}}\label{eq:Fcore}\end{equation}
To find the equilibrium value of $N_{gl}$ at given $R$, the free energy, eq \ref{eq:Fcore}, should be minimized with respect to $N_{gl}$:
\begin{equation}
\frac{dF_{B}}{dN_{gl}}=\frac{\partial F_{B}}{\partial N_{gl}}+
\frac{\partial F_{B}}{\partial R_{gl}}\cdot\frac{\partial R_{gl}}{\partial N_{gl}}=0,\label{eq:FE_min}
\end{equation}
which leads to the following equation
\begin{equation}
\frac{3p^{2}}{32\pi^{2}\varphi^{2}R_{gl}^{4}a^{2}}+\gamma\cdot\frac{2}{\varphi R_{gl}}-\frac{3p(R-R_{gl})}{4\pi\varphi R_{gl}^{2}
(N_{B}-N_{gl})a^{2}}+\mu+\frac{3(R-R_{gl})^{2}}{2(N_{B}-N_{gl})^{2}a^{2}}=0.\label{eq:F_min1}
\end{equation}
Solution of eq \ref{eq:F_min1} should give the equilibrium number of
monomer units (in a star arm), $N_{gl}^{eq}$, that remain in the
globular core as function of the imposed star deformation $R$. 
%Unfortunately,
Eq \ref{eq:F_min1} can be resolved analytically with respect to 
%$N_{gl}$,
%but its solution in the {}``inverse form'' 
$R=R(N_{gl})$.
%can be
%easily found since \ref{eq:F_min1} is none other than a quadratic
%equation in $(R-R_{gl})${[}or in $x:=(R-R_{gl})/(N_{B}-N_{gl})${]}.
The solution is as follows:
\begin{equation}
R=R_{gl}+(N_{B}-N_{gl})a\left[\frac{p}{4\pi\varphi R_{gl}^{2}a} + \sqrt{-\frac{2}{3}\mu-\frac{4\gamma}{3\varphi R_{gl}}} \right],
\label{eq:R_vs_Ngl}
\end{equation}
where $R_{gl}=R_{gl}(N_{gl})$ is given by eq \ref{eq:Rgl}. 
% For $p=1$ it is reduced to. Eq. (40) in \cite{Polotsky:2010}.
\begin{figure}[t]
\includegraphics[width=8cm]{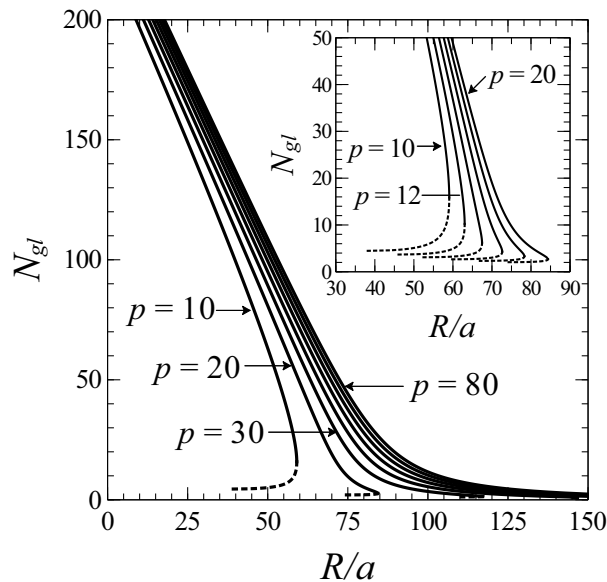}
\caption{\label{fig:Ngl_vs_R} Number of monomers in the star core $N_{gl}$ as function of the star deformation $R$ calculated using eq \ref{eq:R_vs_Ngl}
for $\chi=1$, $N_{B}=200$ and various values of $p=10,\,20,\,30,\, \dots, 80$ in the main figure and $p=10,\,12,\,14,\, \dots, 20$  in the inset. Dashed parts of the curves correspond to unphysical solution.}
\end{figure}
Let us consider eqs \ref{eq:R_vs_Ngl} and \ref{eq:Rgl} as an implicit dependence of the number of monomer units in the core, $N_{gl}$, on the given value of the star extension $R$. 
Figure \ref{fig:Ngl_vs_R} shows a family of $N_{gl}$ vs $R$ curves for $N_{B}=200$, $\chi=1$, and various number of arms $p$. Solid lines show solutions corresponding to $N_{gl}$ that decreases with increasing extension $R$ whereas dashed lines show unphysical situation (simultaneous increase of the deformation $R$ and of the amount of the polymer
condensed in the globular core) and should be therefore excluded from our consideration. The point where $dR/dN_{gl}=0$ represents a spinodal point where the two-phase ``octopus'' state (Figure \ref{fig:B_Star}~b) looses its stability. It corresponds to the (i) largest possible extension and (ii) smallest possible core size for the locally stable microsegregated deformed star. 
%
% With the following increase in $R$ the system passes jumpwise into completely unfolded stretched state (\ref{fig:B_Star}~c). 
%
It can bee seen that for moderate values of $p$ the smallest number of monomer units in the core $N_{gl,\, min} >1$, 
for instance at $p=20$, $N_{gl,\, min} \approx 2.7$. As $p$ grows $N_{gl,\, min}$ decreases and assumes values less than unity.

\begin{figure}[t]
\includegraphics[width=8cm]{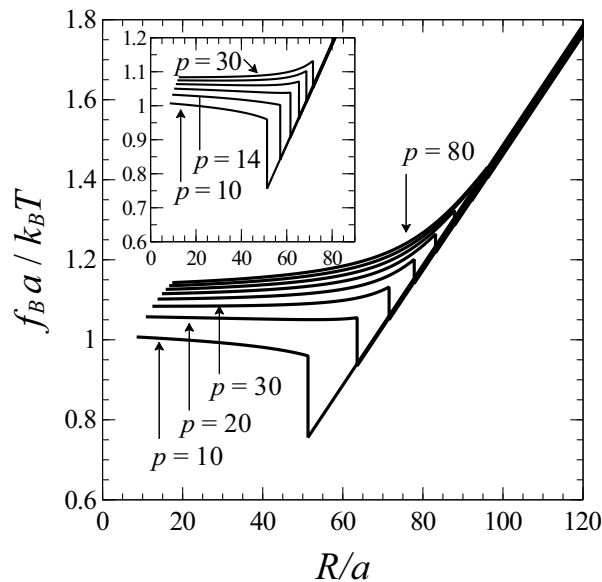}
\caption{\label{fig:force-extension} Equilibrium force-extension curves for
$\chi=1$, $N_B=200$ and various values of  $p=10,\,20,\,30,\, \dots, 80$ in the main figure and $p=10,\,14,\,18,\, \dots, 30$ in the inset.}
\end{figure}

By substituting eq \ref{eq:R_vs_Ngl} into eq \ref{eq:Fcore} one obtains
the equilibrium free energy of the segregated deformed star (Figure \ref{fig:B_Star}~b) as function
of $N_{gl}$; the latter expression together with eq \ref{eq:R_vs_Ngl}
provides a parametric dependence of $F_{B}(R)$ . This
free energy should be compared with that of completely unfolded star
which has no globular core (Figure \ref{fig:B_Star}~c), 
\begin{equation}
\frac{F_{B}^{(unfolded)}}{k_{B}T}=\frac{3(R-r_{0})^{2}}{2N_Ba^{2}},
\label{eq:Funfcore}
\end{equation}
to decide which state of the star (partly or completely unfolded)
is thermodynamically stable.
The comparison of the free energies of the star with partially and completely unfolded arms enables us to localize the
first-order transition point between these two conformations and to calculate the magnitude of the jump-down in the reaction force.
As follows from results presented in Figure \ref{fig:force-extension}, an increase in the number of arms leads to the shift of the transition point 
towards larger values of deformation $R$ and to a decrease in the magnitude of the force jump.
% In order to find the transition point between partially and completely unfolded conformations, the free energy,
% Eq \ref{eq:Funfcore} has to be compared to that corresponding to the partially unfolded conformation and given by Eq...
The reaction force $\mathfrak{f}_{B}$ per arm of the partially unfolded core is:
\begin{equation}
\mathfrak{f}_{B}(R)=\frac{dF_{B}}{dR}=\frac{\partial F_{B}}{\partial R}+\frac{\partial F_{B}}{\partial N_{gl}}\cdot\frac{\partial N_{gl}}{\partial R}=\frac{3(R-R_{gl})}{(N_{B}-N_{gl})a^{2}}\cdot k_{B}T\label{eq:forcecore}\end{equation}
Using eq \ref{eq:R_vs_Ngl} we can express $\mathfrak{f}_{B}$ as a
function of $R_{gl}$:
\begin{equation}
\frac{\mathfrak{f}_{B}(R_{gl})}{k_{B}T}=\frac{3}{a}\left[\frac{p}{4\pi\varphi R_{gl}^{2}a}+\sqrt{-\frac{2}{3}\mu-\frac{4\gamma}{3\varphi R_{gl}}}\right]\label{eq:forcecore_vs_Ngl}\end{equation}
The reaction force per arm of completely unfolded/unraveled star is 
\begin{equation}
\mathfrak{f}_{B}^{(unfolded)}(R)=\frac{\partial F_{B}^{(unfolded)}}{\partial R}=\frac{3(R-r_{0})}{N_{B}a^{2}}\cdot k_{B}T,
\label{eq:forceunfcore}
\end{equation}
Figure \ref{fig:force-extension} shows equilibrium force-extension curves
calculated for the set of parameters ($N=200$, $\chi=1$) similar
to that in Figure \ref{fig:Ngl_vs_R}. 
%The curves start at $R=R_{0}$ with a nonzero force. 
It can be seen that if the number of star arms is
small ($p=10$ or 14 in Figure \ref{fig:force-extension}), the force-extension curve looks very similar to that for a
single chain \cite{Polotsky:2009,Polotsky:2010}.
There is, however, a tiny difference in the initial part of the force-deformation curves:
For the star the curve starts at $R=R_{0}$ with a non-zero force. This is because
in the present work we do not consider an initial deformation regime
that could lead to a distortion of the spherical symmetry of the
collapsed star or to a decrease in the mean polymer concentration $\varphi$
within the star. 
Hence, a pulling force which is less than the threshold force that gives rise to the microphase 
segregation does not affect the structure of collapsed star.
%As a consequence, a non-zero force exceeding a certain
%threshold should be applied to the ends of the arms to initiate the microphase
%segregation within the B-core. 
The value of this threshold force can
be easily obtained from eq \ref{eq:forcecore_vs_Ngl} by setting $N_{gl}=N_{B}$ and $R_{gl}=R_{0}$ :
\begin{equation}
\frac{\mathfrak{f}_{B,min}}{k_{B}T}=\frac{3}{a}\left[\frac{p}{4\pi\varphi R_{0}^{2}a}+
\sqrt{-\frac{2}{3}\mu-\frac{4\gamma}{3\varphi R_{0}}}\right]
\label{eq:threshold_force}
\end{equation}
One can see that the threshold force grows with increasing number of arms in the star $p$. 
% This is related mainly to increasing $R_0$ and the corresponding weakening of the influence of the surface energy penalty in the collapsed core. 
Similar effect is observed for a globule formed by a linear macromolecule: the value of the force corresponding the onset of the microphase segregation in the deformed globule grows as the degree of polymerization increases. In the star case the overall molecular weight is $Np$ and this causes the effect.
In the phase segregation regime, the force for the star with $p=10$ weakly decreases as a function of the
arm extension, in the ``unraveling'' transition point it drops
down and then grows as $3(R-r_{0})/(N_Ba^{2})\simeq3R/(N_Ba^{2})$ for
the unfolded star. However, with an increase in $p$ the character
of the force-extension curves changes: the slope of the quasi-plateau
on the $\mathfrak{f}_{B}(R)$ curve becomes less negative and can
even be positive. To estimate the slope of the force-extension curve,
the derivative $d\mathfrak{f}_{B}/dR$ should be calculated. Since
the reaction force is given in eq \ref{eq:forcecore_vs_Ngl} as a function
of $R_{gl}$, it is expressed as follows
\begin{equation}
\frac{d\mathfrak{f}_{B}}{dR}=\frac{\partial\mathfrak{f}_{B}}{\partial R_{gl}}\cdot\frac{\partial R_{gl}}{\partial N_{gl}}\cdot\frac{dN_{gl}}{dR}\end{equation}
Since $\partial R_{gl}/\partial N_{gl}>0$ 
%(as the number of monomer
%units in the core increases, the core radius increases too; this can
%be also formally checked from \ref{eq:Rgl}), 
and $dN_{gl}/dR<0$,
%(the number of monomer units in the core decreases as we stretch the
%star arms; this can be also seen in \ref{fig:Ngl_vs_R}), 
the sign of $d\mathfrak{f}_B/dR$ is opposite to that of $\partial\mathfrak{f}_{B}/\partial R_{gl}$.
%determines the slope
%of the quasi-plateau. Namely, the reaction force in the quasi-plateau
%regime decreases with increasing deformation if $\partial\mathfrak{f}_{B}/\partial R_{gl}>0$,
%whereas an increasing force corresponds to $\partial\mathfrak{f}_{B}/\partial R_{gl}<0$.
Expression for $\partial\mathfrak{f}_{B}/\partial R_{gl}$ can be
easily found:
\begin{equation}
\frac{\partial}{\partial R_{gl}}\left(\frac{\mathfrak{f}_{B}}{k_{B}T}\right)=\frac{3}{a}\left[-\frac{p}{2\pi\varphi R_{gl}^{3}a}+\frac{2\gamma}{3\varphi R_{gl}^{2}\sqrt{-\frac{2}{3}\mu-\frac{4\gamma}{3\varphi R_{gl}}}}\right]\label{eq:dforce_dRgl}\end{equation}
We see that at small $p$ the second positive term in square brackets
in eq \ref{eq:dforce_dRgl} related to the globule's surface tension dominates, then $d\mathfrak{f}_B/dR <0$, and the quasi-plateau has a negative
slope, whereas for large $p$ the first negative term dominates and
this corresponds to the increasing quasi-plateau. 
% The first negative term in square brackets in \ref{eq:dforce_dRgl} growing with $p$ is an additional contribution 
% characteristic for the star related to the distribution of monomer units pulled from the core between $p$ star arms.
The threshold value
of $p=p^{*}$ at which the negative-to-positive change in the quasi-plateau
slope occurs can be easily found by considering the initial part of
the force-extension curve where $N_{gl}=N_B$, $R_{gl}=R_{0}$. If $p=p^{*}$
corresponds to $\partial f_{B}/\partial R_{gl}=0$, this leads to
the following equation
\begin{equation}
-\frac{p}{2\pi\varphi R_{0}a}+\frac{2\gamma}{3\varphi\sqrt{-\frac{2}{3}\mu-\frac{4\gamma}{3\varphi R_{0}}}}=0\label{eq:pstar_eq}\end{equation}
We can find an explicit solution of the latter equation if we take
$R_{0}\simeq(3N_Bp/4\pi\varphi)^{1/3}$ and $\sqrt{-\frac{2}{3}\mu-\frac{4\gamma}{3\varphi R_{0}}}\simeq\sqrt{-\frac{2}{3}\mu}$.
With these simplifications we obtain
\begin{equation}
p^{*}=4\pi\left(\frac{3N_B}{\varphi a^{3}}\right)^{1/2}\frac{(\gamma a^{2})^{3/2}}{(-6\mu)^{3/4}}.\label{eq:pstar}\end{equation}
Hence, $p^{*}$ increases upon an increase in the solvent strength for the B-block.
For the case $N_B=200$, $\chi=1$ presented in Figure \ref{fig:Ngl_vs_R}
and Figure \ref{fig:force-extension}, eq \ref{eq:pstar} gives $p^{*}\approx25$
which is in accordance with the results presented in the inset in
Figure \ref{fig:force-extension}. Note also that according to eq \ref{eq:pstar}
$p^{*}$ scales as a square root of arm length $N_B$.
At the same time, as $p$ increases, the drop in the reaction force
that accompanies the unraveling transition decreases. This
is in accordance with the conclusion made for a deformed single globule where the force jump at the tadpole - uniformly stretched chain transition decreases as the
overall chain length $N$ increases. 
\section{Conformational transition in (AB)$_{p}$ star block copolymer}
%\subsection{Free energy of the (AB)$_{p}$ star as a function of the core radius}
%Having obtained the expression for the core and the corona free energies,
%we are in the point to to calculate the free energy of the (AB)$_{p}$
%core-shell star as a whole. If we assume that the radius of the B-domain
%is equal to $R$ then the star free energy is given by \ref{eq:Fstar}.
\begin{figure}[t]
\includegraphics[width=6cm]{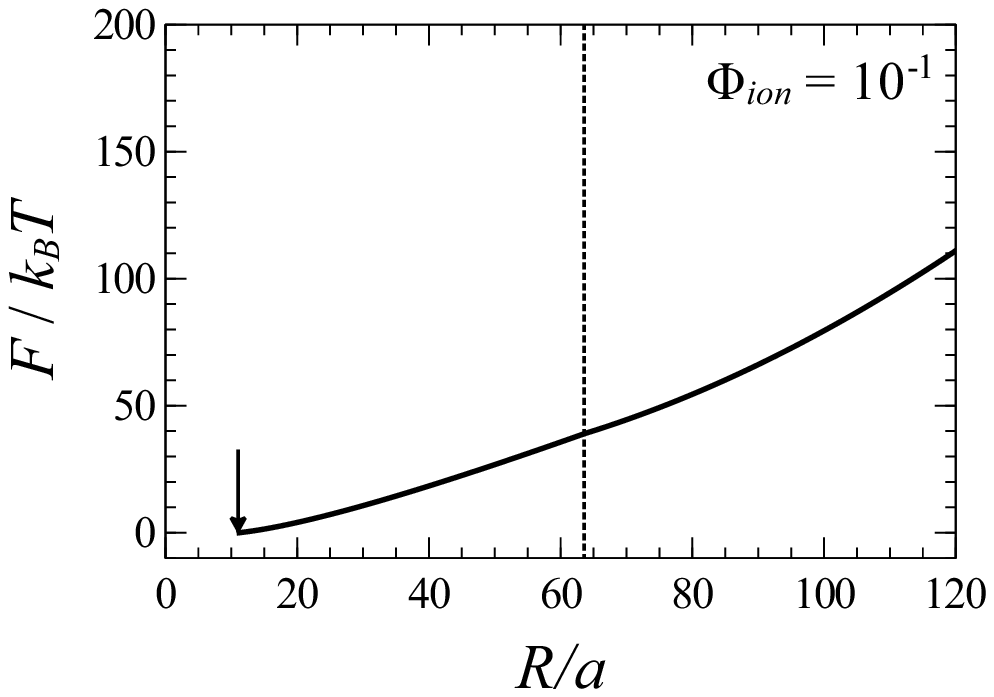}
\includegraphics[width=6cm]{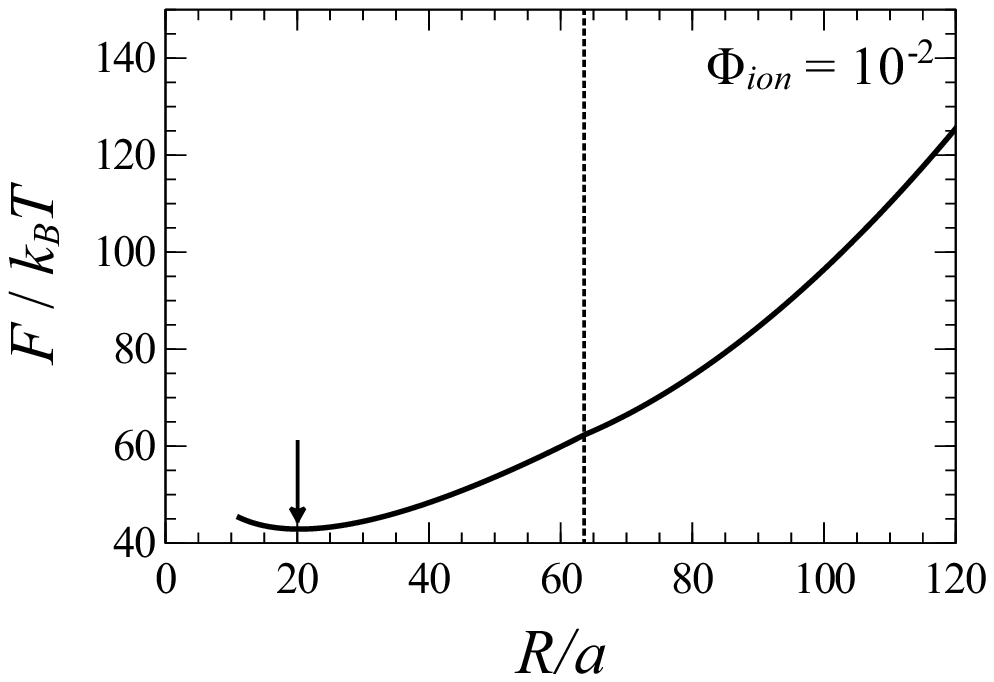}
\includegraphics[width=6cm]{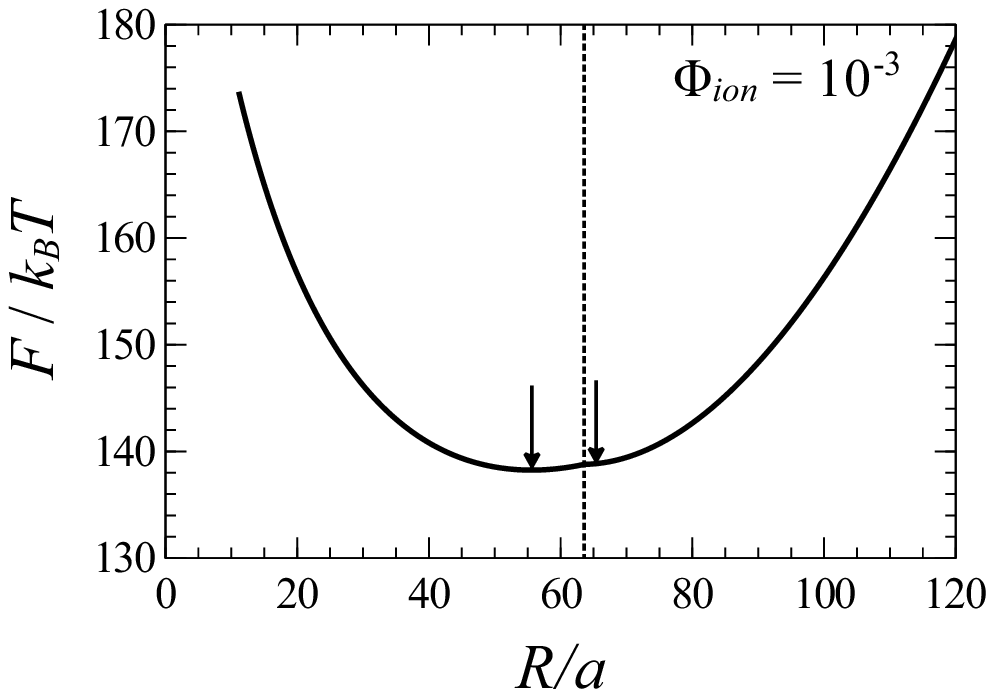}
\includegraphics[width=6cm]{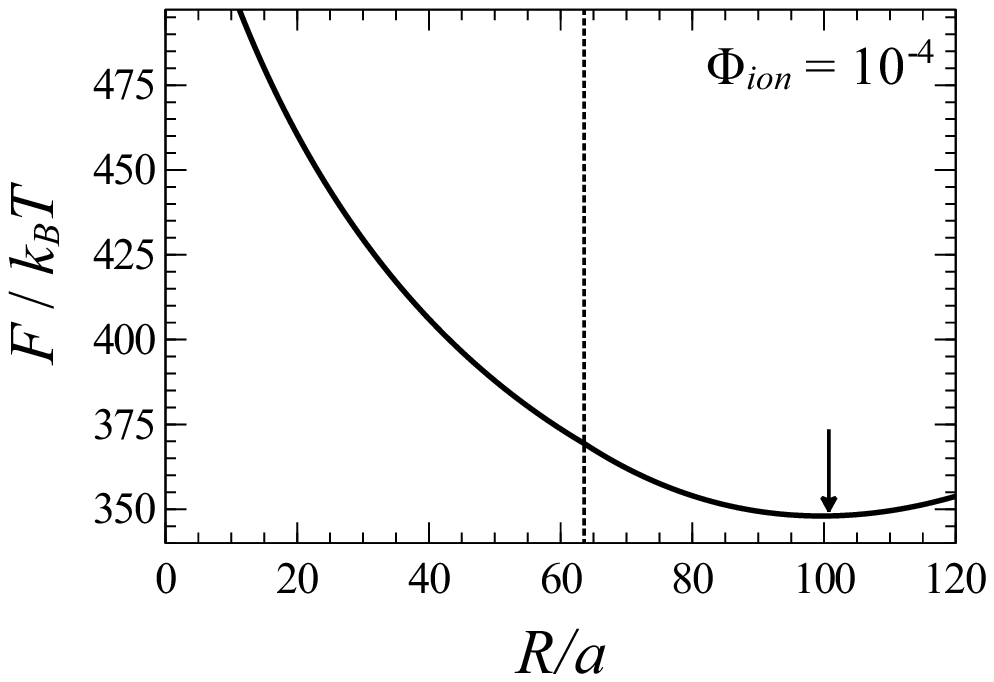}
\caption{\label{fig:Fstar}Free energy of (AB)$_{p}$ core-shell star with $p=20$, $N_{A}=1000$,
$N_{B}=200$, $v_{A}=a^3/6$, $\chi_{B}=1$, $\alpha_{b}=0.5$ and
different values of $\phi_{ion}.$ Arrows show the position of local or boundary
minima, dashed line separates regions of partially and completely
unfolded core.}
\end{figure}
\begin{figure}[t]
\includegraphics[width=8cm]{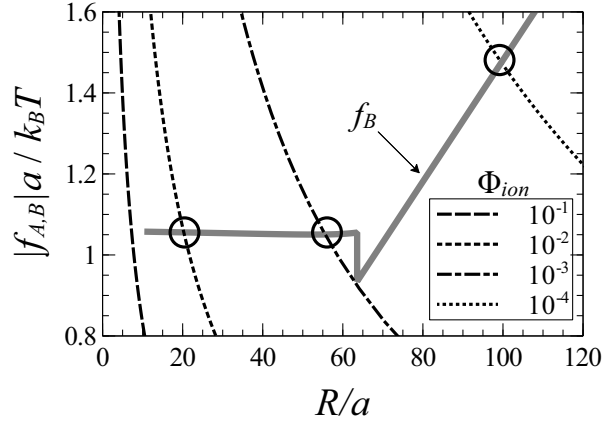}
\caption{\label{fig:forces_AB} Core reaction force (gray curve) and corona pulling force (black curves) in (AB)$_{p}$ core-shell star with $p=20$, $N_{A}=1000$,
$N_{B}=200$, $v_{A}=a^3/6$, $\chi_{B}=1$, $\alpha_{b}=0.5$ and
different values of $\phi_{ion}.$ Circles show the position of equilibrium between the core and the corona.}
\end{figure}
In the preceding sections the expression for the core and the corona free energies and corresponding forces were obtained. Consider now the (AB)$_{p}$
core-shell star as a whole. Figure \ref{fig:Fstar} presents as an example the free energy of the star $F(R)=F_A(R)+F_B(R)$ as a function
of the radial position $R$ of the A-B junction points at different salt concentrations ($F_A(R)$ and $F_B(R)$ were calculated using expressions obtained in the previous sections). An equilibrium conformation of the star corresponds to a (global) free energy minimum. The latter is equivalent to the core and corona force balance condition, eq \ref{force_balance}, i.e. alternatively we can consider force $\mathfrak{f}$ vs radius $R$ dependences. Such dependences corresponding to the case presented in Figure \ref{fig:Fstar} are shown in Figure \ref{fig:forces_AB}: for the core, there is a single curve starting at $r=R_0$ while for the corona, there is a family of curves corresponds to different salt concentrations. Figure \ref{fig:forces_AB} can also be considered as a (partial) superposition of Figure \ref{fig:Fcorona}~b and Figure \ref{fig:force-extension}.

At relatively high salt concentration, $\phi_{ion}=10^{-1}$ the electrostatic repulsions
are weak, and, as a result, the pulling force from the corona with the inner radius $R_0$, eq \ref{eq:R_0}, is smaller than the threshold force, eq \ref{eq:threshold_force}, necessary for unfolding of the core, We see that in this regime the free energy has one boundary minimum corresponding to a unimolecular micelle with collapsed core ($R=R_{0}=(3Np/4\pi\varphi)^{1/3}$). Correspondingly, the $|\mathfrak{f}_A(R)|$ curve does not cross $\mathfrak{f}_B(R)$ dependence, as it can be seen in Figure \ref{fig:forces_AB}.

At lower salt concentration $\phi_{ion}=10^{-2}$, which is below certain salt concentration threshold, $|\mathfrak{f}_A(R)|$ curve crosses $\mathfrak{f}_B(R)$ curve in the microphase segregation (``octopus'') regime at $R > R_0$. This point corresponds to the local (non-boundary) minimum of the free energy $F(R)$. A similar picture is observed at $\phi_{ion}=10^{-3}$, the equilibrium radius of the B-domain, R, is shifted towards larger values. At $\phi_{ion}=10^{-4}$ the free energy minimum and the point of intersection of $|\mathfrak{f}_A(R)|$ and $\mathfrak{f}_B(R)$ curves correspond to the conformation with completely unfolded core.
% at $\phi_{ion}=10^{-4}$.
%a local minimum in the 
%free energy $F_{star}(R)$ as a function of $R$ appears, this minimum is found at $R\geq R_0$ 
%and is shifted 
%towards larger values of $R$ upon further decrease in the salt concentration. 
%This minimum in the free energy corresponds to a microphase segregated states at 
%$\phi_{ion}=10^{-2},10^{-3}$ or
%to the conformation with completely unfolded core at $\phi_{ion}=10^{-4}$ (see \ref{fig:Fstar}).
Remarkably, close to the transition point between the conformations with partially and fully unfolded core, the free
energy has two minima (this can be seen on curves for $\phi_{ion}=10^{-3}$
in Figure \ref{fig:Fstar}; in Figure \ref{fig:forces_AB} two points of intersection are clearly seen). Besides the main minimum corresponding to the octopus-like conformation, the second metastable minimum related to completely unfolded core appears. 
The appearance of the two minima in the free energy implies co-existence of 
the stars with the partially and
completely unfolded core domain and 
signifies quasi-first order 
character of the unfolding transition: 
Upon a decrease in $\phi_{ion}$ the system "jumps" from
the "left-hand-side" minimum (corresponding to the microphase-segregated partially unfolded core) to the "right-hand side"
minimum (corresponding to the star with completely unfolded core domain).
%According to the results presented in \ref{fig:Fstar} (i.e. 
For the set of parameters used here,
the co-existence between microphasely-segregated and fully unfolded core
occurs at $R\approx61$, as it is shown in by a vertical dashed line in Figure \ref{fig:Fstar}. 
%
% {\bf Do we need here a figure similar to Figure 2, but with the A blocks upending the B-blocks?}
%
\begin{figure}[t]
\includegraphics[width=8cm]{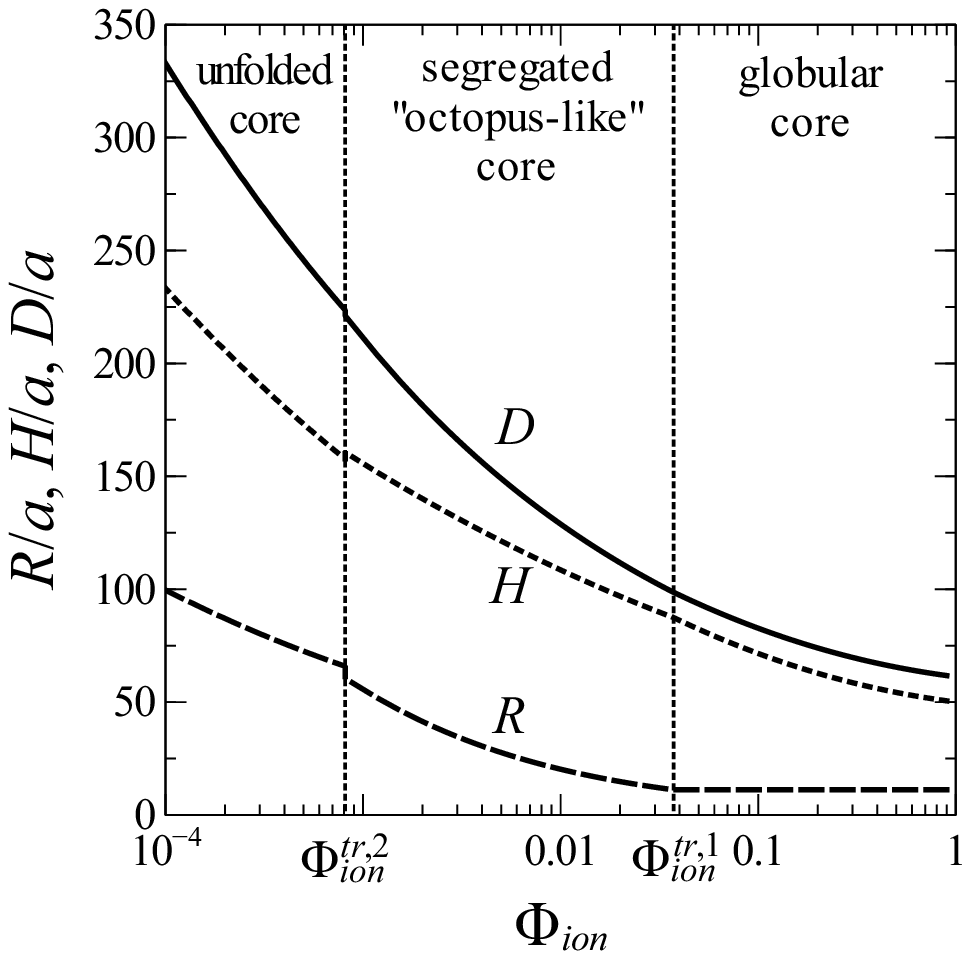}
\caption{\label{fig:R_vs_phi} Core radius $R$ (dashed line), corona thickness $H$ (dotted line) and overall star radius $D=R+H$ (solid line) as functions of ionic strength. Parameters of the (AB)$_{p}$ star: $p=20$, $N_{A}=1000$, $N_{B}=200$, $v_{A}=a^3/6$, $\chi_{B}=1$, $\alpha_{b}=0.5$. Vertical dotted lines separate different star regimes.}
\end{figure}
Figure \ref{fig:R_vs_phi} 
%presents the result of minimization of $F_{star}$ with respect to $R$ and 
shows equilibrium position $R$ of the A-B junction points, the
overall star radius $D$ (which is, equivalently, the outermost corona radius), and the corona thickness
$H=D-R$ as functions of salt concentration $\phi_{ion}$. 
As we have shown, with a
decrease in $\phi_{ion}$ the star undergoes a sequence of intramolecular
conformational transitions: when salt concentration is large, repulsive
interactions in the corona are weak and the core keeps
the spherical collapsed shape with the radius $R_{0}.$
The corona thickness
and overall star size increase in this regime upon a decrease in salt concentration
due to progressively increasing differential osmotic pressure inside the corona.
Simultaneously, the pulling force applied to the ends of the B-blocks localized at the surface of the
core domain increases. 
At certain
threshold salt concentration $\mathfrak{f}_{A}(R_0)$ becomes equal to the critical
minimum force $\mathfrak{f}_{B,min}$ given by eq \ref{eq:threshold_force}.
% \ref{eq:forceunfcore}.
This point
corresponds to the onset of the microphase segregation within the core,
and appearance of the stretched ``legs'. 
From this
point on, a decrease in salt concentration leads to a progressive unfolding of the core,
an increase in the length of the ``legs'' and, consequently, to an 
increase in $R$, $D$, and $H$, which is interrupted by abrupt transition
from partially to completely unfolded core conformation.
In the transition
point the core size $R$ increases jumpwise and this causes a
simultaneous jumpwise decrease of the corona thickness $H$. This 
drop in the corona thickness occurs because of increasing distance between the 
A-B junction points and decreasing crowding of the A-blocks in the coronal domain.
However, as a
result of these two opposite tendencies the overall (outermost) size of the star $D$ exhibits a small jump up. 
In the
law salt regime where the core is completely unfolded, $R$, $D$,
and $H$ grow again progressively as $\phi_{ion}$ decreases.
\begin{figure}[t]
\includegraphics[width=8cm]{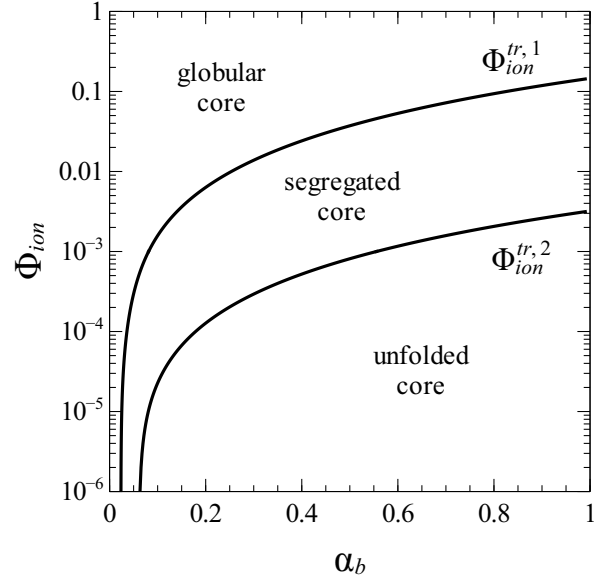} 
\caption{\label{fig:phd_alpha}Diagram of states of amphiphilic (AB)$_{p}$
core-shell star with $p=20$, $N_{A}=1000$, $N_{B}=200$, $v_{A}=a^3/6$,
$\chi_{B}=1$ on $(\alpha_{b};\phi_{ion})$ plane. $\phi_{ion}^{tr,1}$ and $\phi_{ion}^{tr,2}$ are the ionic strengths corresponding to the onset of microphase segregation in the core and to complete core unfolding, respectively.}
\end{figure}

\begin{figure}[t]
\includegraphics[width=8cm]{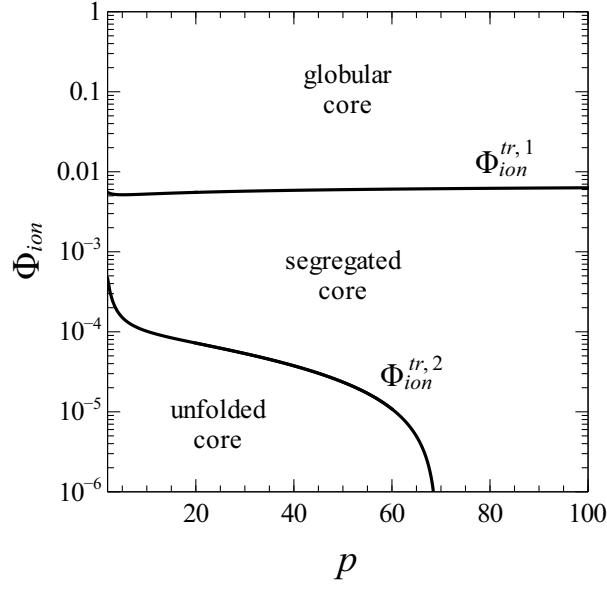}
\caption{\label{fig:phd_p} Diagram of states of amphiphilic (AB)$_{p}$ core-shell
star with $N_{A}=500$, $N_{B}=200$, $v_{A}=1/6$, $\chi_{B}=1$,
$\alpha_{b}=0.2$ on $(p;\phi_{ion})$ plane. $\phi_{ion}^{tr,1}$ and $\phi_{ion}^{tr,2}$ are the ionic strengths corresponding to the onset of microphase segregation in the core and to complete core unfolding, respectively.}
\end{figure}

We see that a progressive decrease in $\phi_{ion}$ causes, in a sequence,
two intermolecular conformational transitions: 
The first one corresponds
to the continuous onset of the microphase segregation within the core
whereas the second one is the 
% first order 
jumpwise transition corresponding
to the complete unfolding of the core (vanishing of the globular phase).
%It is not a hard task to find 
We denote the salt concentrations at which these transitions occur as $\phi_{ion}^{tr,1}$ and $\phi_{ion}^{tr,2}$,
respectively. Their values are determined by the parameters of the (AB)$_p$ core-shell star ($N_A$, $N_B$, $p$, $\alpha_b$, $v_A$, $\chi_B$), therefore, a multidimensional diagram of states can be constructed. Figure \ref{fig:phd_alpha} and Figure \ref{fig:phd_p} represent two-dimensional sections of the such a diagram of states in coordinates $(\alpha_b,\, \phi_{ion})$ and $(p,\, \phi_{ion})$, respectively.
%The dependences
%of $\Phi_{ion}^{tr,1}$ and $\Phi_{ion}^{tr,2}$ on the degree of ionization $\alpha_{b}$ of the A-blocks 
%correspond to the boundary lines between different regions in the diagram of states shown in \ref{fig:phd_alpha}. 
%The dimensions of the (AB)$_{p}$ star vary continuously upon crossing the $\Phi_{ion}^{tr,1}$ boundary, but exhibit jump-wise variation at the 
%$\Phi_{ion}^{tr,2}$ boundary.
As follows from the analysis of the first diagram, Figure \ref{fig:phd_alpha}, if the fraction
of charged monomers $\alpha_b$ in the corona block A is small (below a certain
threshold value $\alpha_{b,min}$), the electrostatic interactions in the corona are not
strong enough to induce the unfolding of the collapsed core even at vanishing ionic strength of the solution.
The value of $\alpha_{b,min}$ is controlled by the hydrophobicity of the core forming blocks and can be expressed as
\begin{equation}
\alpha_{b,min}\cong\biggl(\frac{a\mathfrak{f}_{B,min}}{k_{B}T}\biggr)^{1/2}
\end{equation}
where $\mathfrak{f}_{B,min}$ is given by eq \ref{eq:threshold_force}.
Moreover, if $\alpha_b$ is only slightly larger than $\alpha_{b,min}$, a decrease in the ionic strength stabilizes the microphase-segregated
state of the core, whereas the complete core unfolding does not occur even in the salt-free solution.

The diagram of states in $(\phi_{ion},\, p)$
coordinates, Figure \ref{fig:phd_p}, allows us to analyze the influence of the number
of the star arms $p$ on the character of the conformational transitions related to the core unfolding. 
The diagram presented in
Figure \ref{fig:phd_p} shows that $\phi_{ion}^{tr,1}$ corresponding to the onset of
the microphase segregation in the core is only a weakly increasing function of $p$.
By contrast, the characteristic salt concentration $\phi_{ion}^{tr,2}$
corresponding to the jumpwise complete core unfolding
is a decreasing function of $p$ which approaches the zero at some finite value of $p$ 
($p\approx70$ for our set of parameters).
In the stars with larger number of arms the core unfolding occurs continuously 
upon a decrease in the salt concentration.
This behavior is related to a decrease of number of monomers in the ``minimal core'' and a decrease in
the magnitude of the jump in the A-B junction points position 
upon transition from the microphase-segregated to completely unfolded state 
as $p$ increases (see
Figure \ref{fig:force-extension} and Section ``Core domain'').
Hence, the microphase-segregated
region on the phase diagram in Figure \ref{fig:phd_p} extends at
large $p$ to low salt concentration but one should realize that at
small $\phi_{ion}$ the core is (almost) completely
unfolded.
\section{SCF modeling}
To check the validity of our theoretical predictions and the correctness
of the underlying model assumptions, we use Scheutjens-Fleer self-consistent
field (SF-SCF) numerical approach to study the structure of the AB
core shell star and conformational rearrangements upon changes in
environmental conditions, i.e., upon change in the ionic strength. 
\subsection{SCF formalism}
The self-consistent field (SCF) method is well-known for the modeling
of inhomogeneous polymer layers. At the basis of the approach is a
mean-field free energy which is expressed as a functional of the volume
fraction profiles (normalized concentrations) and self-consistent
field potentials. The optimization of this free energy leads for Gaussian
polymer chains to the Edwards diffusion (ED) differential equation
\cite{deGennes:1979}, which for an arbitrary coordinate system may
be expressed as
\begin{equation}
\frac{\partial G(\mathbf{r},\mathbf{r}';\, n)}{\partial n}=\frac{a^{2}}{6}\nabla^{2}G(\mathbf{r},\mathbf{r}';\, n)-\frac{u(\mathbf{r})}{k_{B}T}G(\mathbf{r},\mathbf{r}';\, n)\label{eq:EDE}\end{equation}
($k_{B}$ is the Boltzmann constant, $T$ is the absolute temperature,
$a$ is the monomer unit size). 
The Green's function $G(\mathbf{r},\mathbf{r}';\, n)$ used in eq \ref{eq:EDE},
is the statistical weight of a probe chain with the length $n$ having
its ends fixed in the points $\mathbf{r}'$ and \textbf{$\mathbf{r}$}.
The self-consistent potential $u(\mathbf{r})$ represents the surrounding
of the chain and serves as an external field used in the Boltzmann
equation to find the statistical weight for each chain conformation.
Consequently, Greens functions $G(\mathbf{r},\mathbf{r}';\, n)$
that obey eq \ref{eq:EDE} are related to the volume fraction profile
of the polymer $\varphi(\mathbf{r})$ or, in the case of a multicomponent
system, to the set of volume fraction profiles $\varphi_{k}(\mathbf{r})$
where $k=1,2,...$ specifies the component. 
As long as the potential $u(\mathbf{r})$ is local (i.e., there are no long-range
forces) it depends on the local volume fraction $\varphi(\mathbf{r})$.
The electrostatic interactions in the system are taken
into account on the level of the Poisson-Boltzmann approximation: local electrostatic potential
is calculated from the Poisson equation, which accounts for the 
charge density distribution created by all charged species (i.e., charged monomer units
and mobile ions).
This makes up the system of self-consistent field equations which
is solved iteratively: one assumes an initial volume fraction profile
$\varphi(\mathbf{r})$, then computes the potential $u(\mathbf{r})$,
the set of the Greens functions using eq \ref{eq:EDE}, and derives
a new volume fraction profile $\varphi'(\mathbf{r})$. The procedure
is then repeated until the sequence of approximations converges to
a stable solution.
To solve these equations rigorously, it is necessary to introduce
a numerical algorithm. Such numerical scheme invariably involves space
discretization (i.e., the use of a lattice). Here we follow the method
of Scheutjens and Fleer (SF-SCF) \cite{Fleer:1993} who used the segment
size $a$ as the cell size. A mean-field approximation is applied
to a set of lattice sites. This set (often called a lattice layer)
is referred to with a single coordinate \textbf{$\mathbf{r}$}. The
way the sites are organized in layers depends on the symmetry in the
system and must be preassumed. The approach allows for, e.g., volume
fraction gradients between these layers. 
In order to consider an isolated polymer star, it is enough to use
a one-gradient version of SCF algorithm in the spherical coordinate
system, for which \textbf{$\mathbf{r}=r$}. In this case, all volume
fraction profiles as well as other thermodynamic values depend only
on the radial coordinate $r$, while the mean-field approximation
is used along angular coordinates.
More specifically, within our model $p$ AB diblock copolymer chains
are homogeneously pinned at the surface of a small sphere with the
radius $r_{0}=a\left\lceil \sqrt{p/(4\pi)}\right\rceil $, where $\left\lceil x\right\rceil $
denotes is the smallest integer not less than $x$ (the so called
ceiling function).
AB copolymer nature of the chains is taken into account via the set
of Flory-Huggins parameters for monomer-solvent interactions, $\chi_{A}$,
$\chi_{B}$, and monomer-monomer cross-interactions $\chi_{AB}$.
The solvent is assumed to be good (athermal) for A monomers, $\chi_{A}=0$,
and poor for B monomers, $\chi_{B}=1$. We chose $\chi_{AB}=\chi_{B}$
meaning that the incompatibility between A and B monomers is the same
as between B monomers and solvent molecules. 
The block A is a positively charged polyelectrolyte, with quenched
charge, the fraction of charged monomer units is equal to $\alpha_{b}$
(i.e. $\alpha_{b}$ is the fractional charge per A monomer unit). There are two
types of salt ions in the system: co-ions $\mathrm{Na}^{+}$ and counterions
$\mathrm{Cl}^{-}$. 
SCF calculations were performed using the {\em SFBox} program developed at Wageningen Uviversity.
\subsection{Results of SCF modeling}
To make a quantitative comparison with the results of the theory,
we have made SCF calculations for (AB)$_{p}$ copolymer star with the
parameters similar to those used in the previous section (Figures \ref{fig:Fstar}
- \ref{fig:phd_p}), that is, we consider a star with $p=20$ AB-diblock
copolymer arms (for $p=20$ the radius of the grafting sphere is $r_{0}=2$),
$N_{A}=1000$, $N_{B}=200$, fraction of charged A-monomers $\alpha_{b}=0.5$,
and monomer-solvent interaction parameters $\chi_{A}=0$ (which gives
for the second virial coefficient $v_{A}=a^3/6$), and $\chi_{B}=1$.
The Flory-Huggins interaction parameters for the salt ions are taken 
the same as
for the solvent molecules.
The salt volume fraction was varied in the range $10^{-1}\leq\phi_{ion}<10^{-4}$. 
The SF-SCF method has obvious advantages, as compared to the analytical theory, in being
assumption-free: neither strong chain stretching nor equal radial positions of the
junction points and terminal monomer units are pre-assumed.
As a result, the SF-SCF approach provides a more accurate information
about density distribution in the core and in the corona of the star.
On the other hand the use of one-gradient version of the SCF approach ``smears out''
lateral inhomogeneity of the star with microphase-segregated or
unfolded core, Figure \ref{fig:B_Star}.

Since one of our motivations of using the SCF modeling is to check the underlying model assumptions,
let us list the latters. The analytical theory was developed under assumption that 
\begin{enumerate}
\item \label{assump:B} all hydrophobic B-blocks are equally (but possibly not uniformly) stretched so all the 
A-B junction points are located at equal distance $R$ from the center of the star;
\item \label{assump:A} all hydrophilic A-blocks are equally (but not uniformly) stretched so all the A-endpoints are located on the equal distance $D$ from the center of the star;
\item \label{assump:segregation} core and corona (correspondingly B and A blocks) are segregated in space: there is a sharp phantom interface separating core and coronal domains;
\item \label{assump:ms} at strong repulsive interaction in the corona microphase segregation in the core occurs;
\item \label{assump:legs} at strong repulsive interaction in the corona stretched ``legs'' of the microphase-segregates ``octopus'' structure are {\em uniformly} extended;
\item \label{assump:LEA} local electroneutrality approximation for the corona is fulfilled;
\end{enumerate}
\begin{figure}[t]
(a) \includegraphics[width=7cm]{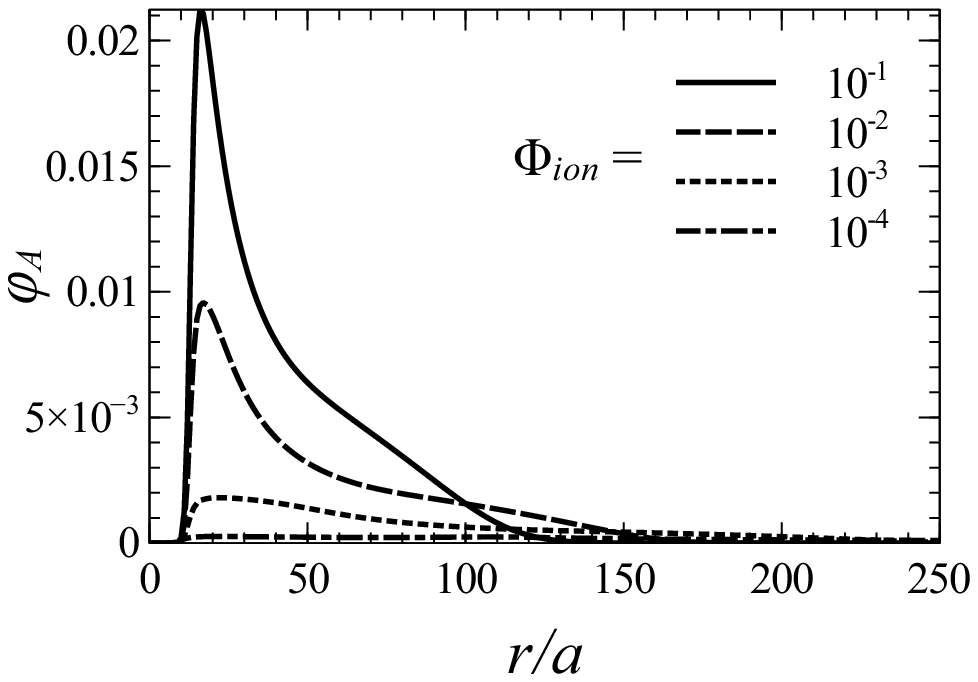} \includegraphics[width=7cm]{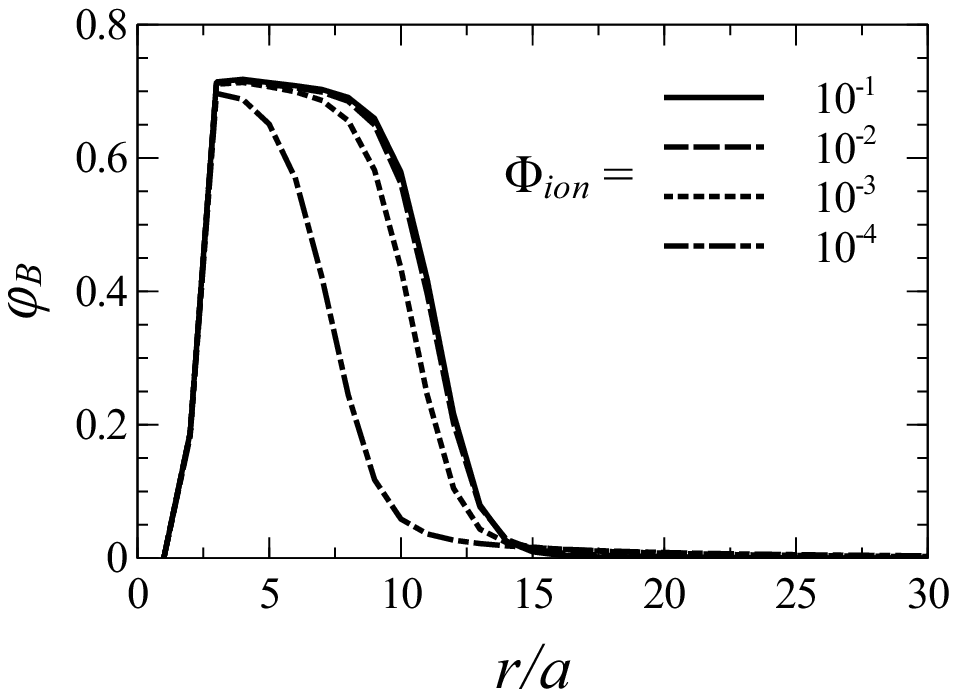} (b)
\caption{\label{fig:SCF_phi}Radial density profiles of A-monomers (a) and
B-monomers (b) in AB-core-shell star with $p=20$, $N_{A}=1000$,
$N_{B}=200$, $\chi_{A}=0$, $\chi_{B}=1$, $\alpha=0.5$ and different
values of $\phi_{ion}$ (shown in the legend) .}
\end{figure}
Figure \ref{fig:SCF_phi} shows evolution of radial density profiles of A-
and B-monomers with decreasing ionic strength. 
%To resolve fine details of the profiles in the range of very low density, the profiles are presented not only in linear but in semi-logarithmic coordinates as well. 
The density profile of the swollen corona (A-units density profile,
Figure \ref{fig:SCF_phi} a) has a sharp maximum at the
interface between A and B domains and then decays with increasing $r$. 
%Remarkably, the concentration of the A-monomer units in the B-domain is negligibly small.
A decrease in the salt concentration $\phi_{ion}$ leads to progressive corona swelling:
the maximum value of $\varphi_{A}$ decreases, the profile broadens
(i.e. the overall AB star size grows, in accordance with Figure \ref{fig:R_vs_phi}). 
The density profile of hydrophobic B-monomer units is qualitatively very different
from its A-counterpart. Figure \ref{fig:SCF_phi}~b shows that at high
salt concentration ($\phi_{ion}$= 0.1; 0.01) the distribution of the B-units inside the core
is nearly uniform,
with a narrow depletion zone in the layer adjacent to the grafting
point and a sharp decay at the edge of the core. With a decrease in
$\phi_{ion}$ , the pulling force from the coronal A-blocks grows and microphase
segregation in the core is clearly seen. 
%(especially on semi-logarithmic plot \ref{fig:SCF_phi} d). 
The profile has two parts: a dense core
and a sparse periphery with the density much less than that of the
core. As $\phi_{ion}$ decreases, the globular core becomes smaller and
%(this is better seen on linear scale plot, \ref{fig:SCF_phi} c),
the low-density peripheral part extends at the expense of the depleted core.
This picture is in accordance with our model assumption 4 about microphase
segregation in the core. 

Another important characteristic
%interesting property 
is the radial distribution of the number of monomers of the type $i$, $n_i(r)$. It is easily obtained from the corresponding radial
density distribution $\varphi_{i}(r)$ by multiplying it by the number of lattice sites in the corresponding layer $L(r)$ (which is the volume of the spherical layer) : $n_{i}(r)=\varphi_{i}(r)\cdot L(r)=\varphi_{i}(r)\cdot4\pi(ar^{2}-a^{2}r+a^{3}/3)$. Number distribution is the best way for expressing and analyzing the distribution of the end monomers or A-B junction points.
\begin{figure}[t]
(a) \includegraphics[width=7cm]{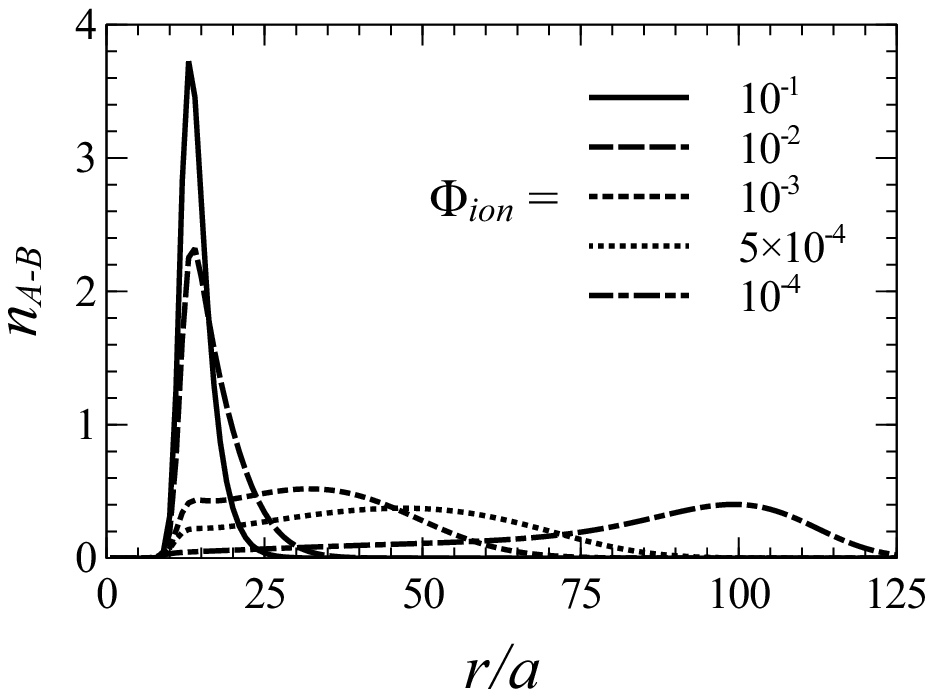} 
\includegraphics[width=7cm]{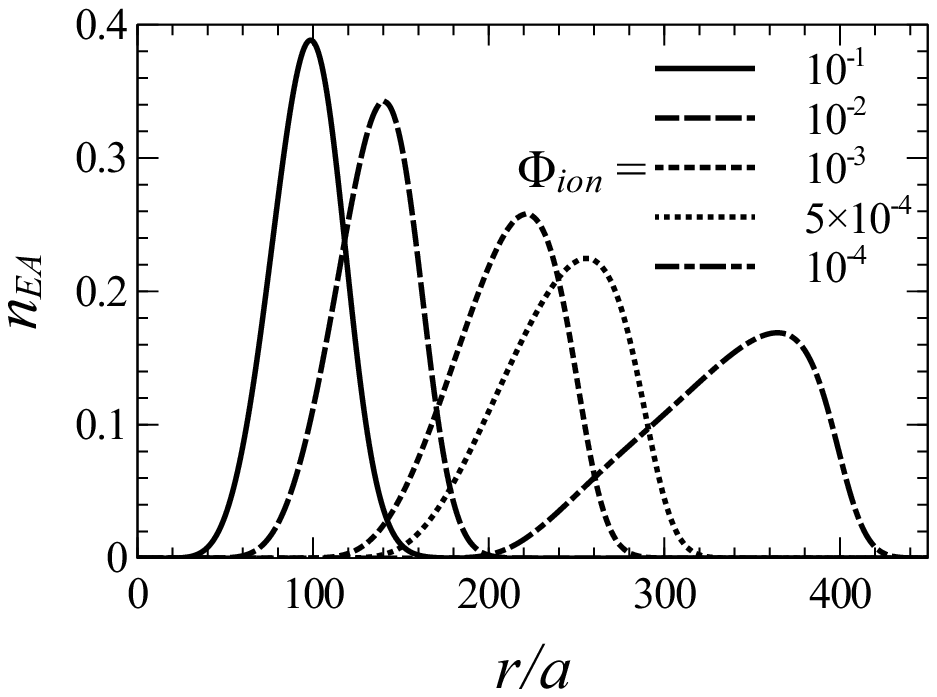} (b)
\caption{\label{fig:SCF_ne}Radial distribution of the number of A-B junction
points (a) and A-endpoints (b) in AB-core-shell star with $p=20$, $N_{A}=1000$, $N_{B}=200$,
$\chi_{A}=0$, $\chi_{B}=1$, $\alpha=0.5$ and different values of
$\phi_{ion}$.}
\end{figure}
The evolution of the distribution of A-B junction points, $n_{J}(r)$,
is shown in Figure \ref{fig:SCF_ne} a. At high salt concentration, when the
pulling force exerted by the corona is not strong enough to provoke microphase segregation
in the core, the distribution is rather narrow with a pronounced peak.
Our calculations demonstrated that the distributions for $\phi_{ion}$= 0.1 and 0.05 (the latter is not shown in  Figure \ref{fig:SCF_ne} a) are indistinguishable, whereas
the following decrease in $\phi_{ion}$ leads to broadening
of the distribution and to progressive shift of the position of the maximum
towards larger $r$. By analyzing the shape of the profiles, we can
conclude that even though distributions are rather broad each profile has one pronounced maximum, and we do not
encounter the situation with two maxima corresponding to partitioning
of the star arms into two groups, i.e., those whose B blocks are completely embedded
in the core and those whose B-blocks are (partially) expelled from the
core and stretched. 
%(even the profiles at $\Phi_{ion}=10^{-3}\,\mathrm{and}\,5\times10^{-4}$
%are hard to be considered as bimodal, corresponding to two populations
%of chains). 
This finding justifies the used by the analytical theory approximation 1 that all
the junction points were placed equidistantly from the center of the
star.
%(hence, the distribution is reduced to a single $\delta$-peak). 

\begin{figure}[t]
\includegraphics[width=7cm]{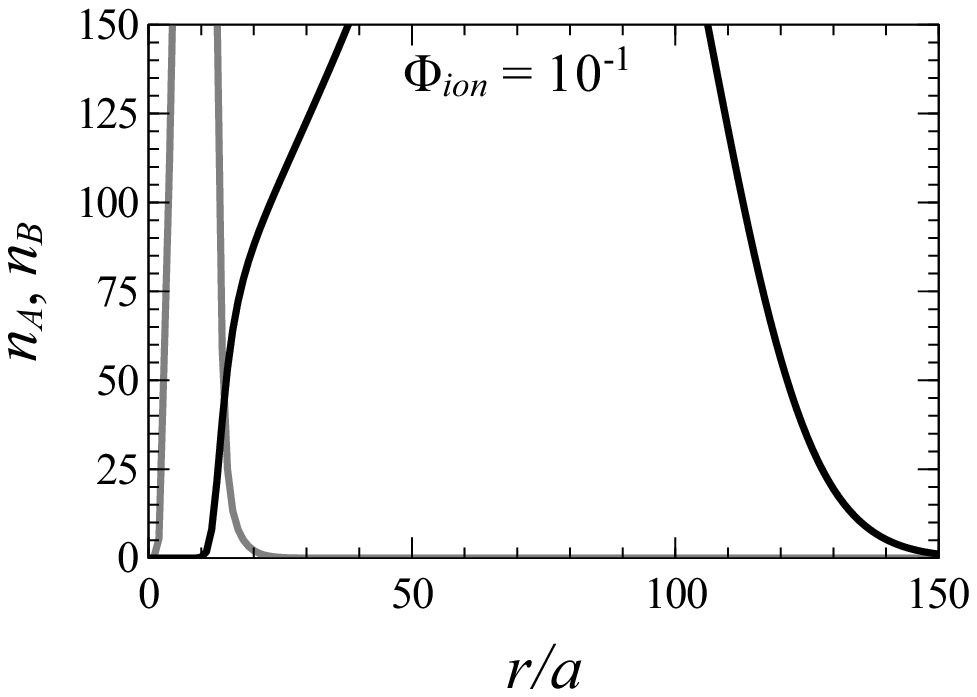}
\includegraphics[width=7cm]{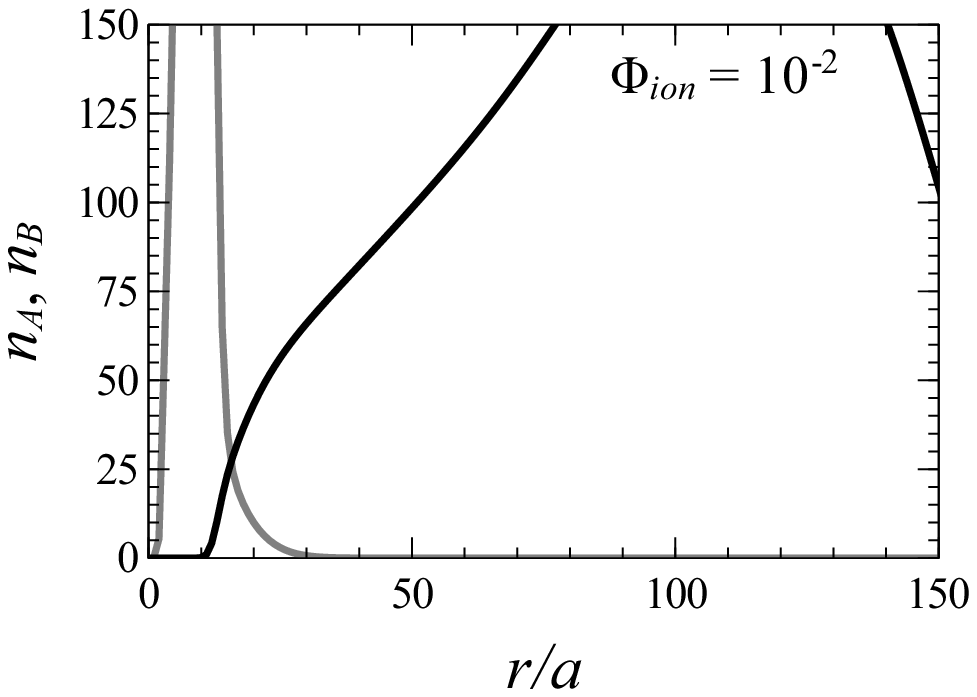}
\includegraphics[width=7cm]{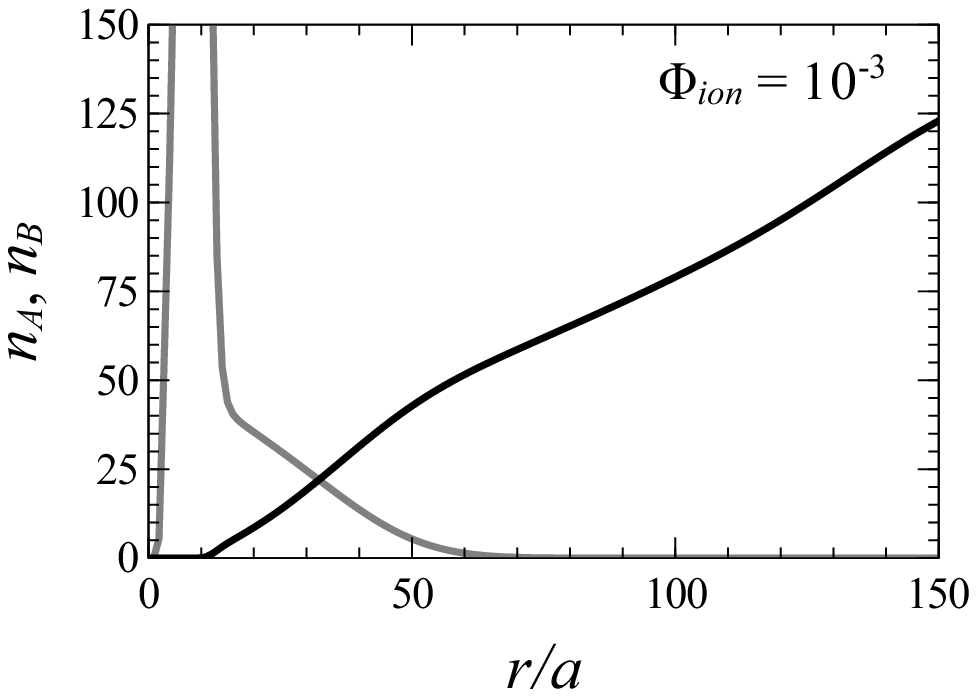}
\includegraphics[width=7cm]{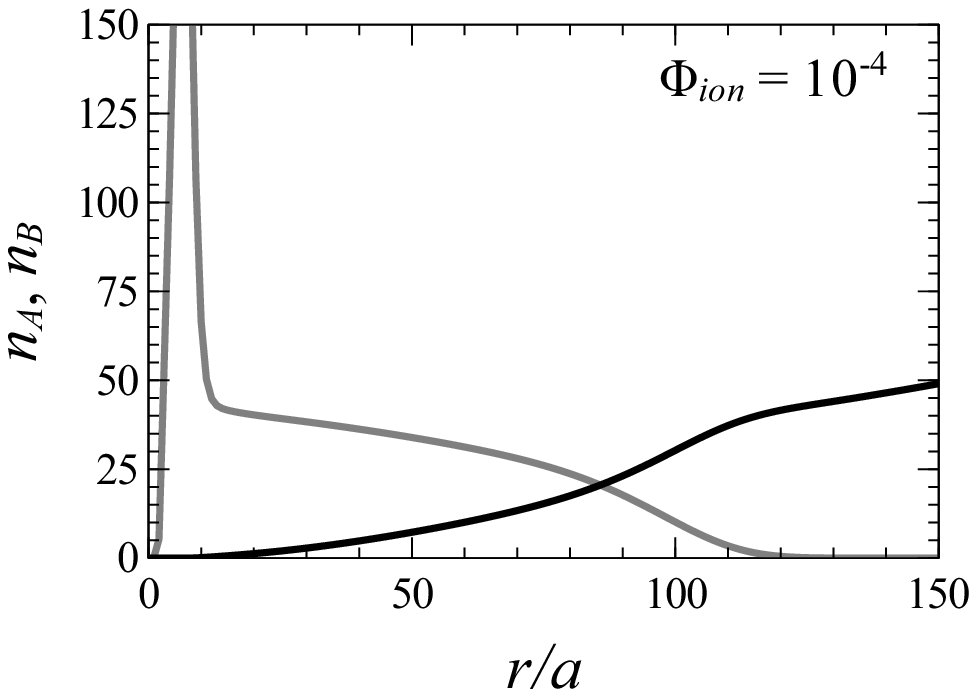}
\caption{\label{fig:SCF_nab}Radial distribution of the number of A- (black curves) and B-monomers (gray curves) in the vicinity of the A/B interface in AB-core-shell star with $p=20$, $N_{A}=1000$,
$N_{B}=200$, $\chi_{A}=0$, $\chi_{B}=1$, $\alpha=0.5$ and different values of $\phi_{ion}$ (shown in each panel).}
\end{figure}

The distribution of ends of corona blocks, Figure \ref{fig:SCF_ne}~b has one pronounced maximum, with a decrease in $\phi_{ion}$ leading to increase of the radius of the A/B interface, $R$, the distribution becomes asymmetric, biased towards the edge of the star. Therefore, the use of assumption 2 is reasonable.
\begin{figure}[t]
\includegraphics[width=7cm]{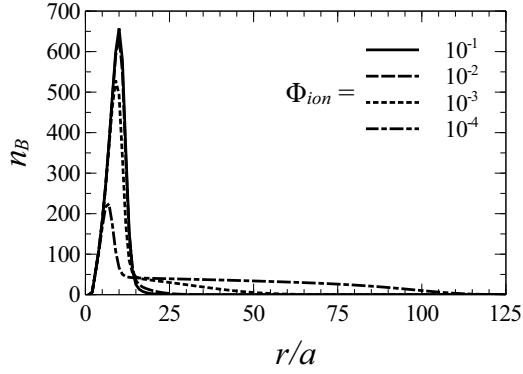}
\caption{\label{fig:SCF_nb}Radial distribution of the number of B monomer units in AB-core-shell star with $p=20$, $N_{A}=1000$, $N_{B}=200$,
$\chi_{A}=0$, $\chi_{B}=1$, $\alpha=0.5$ and different values of
$\phi_{ion}$.}
\end{figure}
Although, as it was noted above, in the one-gradient version of the SF-SCF method we cannot see the stretched ``legs'', 
they can be revealed by plotting the radial profiles of the number of B monomers per spherical layer. 
The $n_B(r)$ distribution allow us to analyze the local tension in the extended part of the core. 
Uniform ``leg'' stretching (according to assumption 5) would correspond to a plateau on the $n_B(z)$ profile. From Figure \ref{fig:SCF_nb} 
we see that in our system this situation is better manifested in the case of very strong corona stretching force (very low $\phi_{ion}$). 
Deviations from the flat plateau shape of the $n_B(z)$ profiles observed in Figure \ref{fig:SCF_nb} (these deviations are also well seen in the 
plots of Figure \ref{fig:SCF_nab}) are due to the finite width of the distribution of the A-B junction points.

Radial distribution of a number of monomer units found at given distance from the center 
is also an appropriate characteristic for estimating the degree of overlap of the core and the corona (hence, we can test the validity of assumption 3). 
These profiles presented together on the same plot are shown in Figure \ref{fig:SCF_nab} (the scale of the plot is chosen for better ``zoom'' of the A/B interfacial region)
We see that $n_A(r)$ and $n_B(r)$ profiles (shown by black and gray lines, 
respectively) demonstrate a small overlap, the ``maximum'' value of the $n_A$ and $n_B$ in the overlap zone corresponds to intersection of the profile and is very close to the value of $p$ (in our case $p=20$). Hence, the fraction of monomers of ``guest'' A (B) component in the ``host'' B-core (A-coronal) domain is small, so we can conclude that the core and the corona are well segregated, even under the conditions of strong stretching of the inner B-blocks, that is the case at low $\phi_{ion}$ values.
\begin{figure}[t]
\includegraphics[width=8cm]{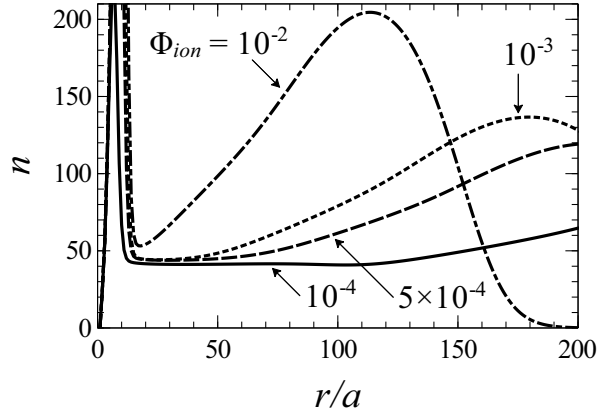}
\caption{\label{fig:SCF_ntot}Radial distribution of total number of monomer units in AB-core-shell star with $p=20$, $N_{A}=1000$, $N_{B}=200$, $\chi_{A}=0$, $\chi_{B}=1$, $\alpha=0.5$ and different values of $\phi_{ion}$ in the vicinity of the A/B interface. }
\end{figure}
By summing up the distributions presented in Figure \ref{fig:SCF_nab} another interesting feature of 
% the system under study 
the star with partially unfolded core is revealed. 
At strong corona pulling forces (low ionic strength), the distribution of the total number of monomer units, 
$n = n_A + n_B$, shown in Figure \ref{fig:SCF_ntot} has a plateau in the vicinity of the A/B boundary. 
This means that the strongly (and homogeneously) stretched ``legs'' include not solely the parts of 
B blocks extracted from the B-core but also parts of A blocks adjoining the ends of the B blocks. 
The uniform stretching of the arms is typical for starlike polyelectrolyte in the 
"osmotic" (dominated by the counterions) regime.
With an increase in $r$ this plateau is followed by the growing part on $n(r)$ profile corresponding 
to the decrease in the local stretching of star arms which is typical for a convex spherical (or cylindrical) polyelectrolyte brush
in the salt-dominance regime~\cite{Birshtein:2008}. 
At moderate and high ionic strength such a plateau is not observed.
\begin{figure}[t]
\includegraphics[width=8cm]{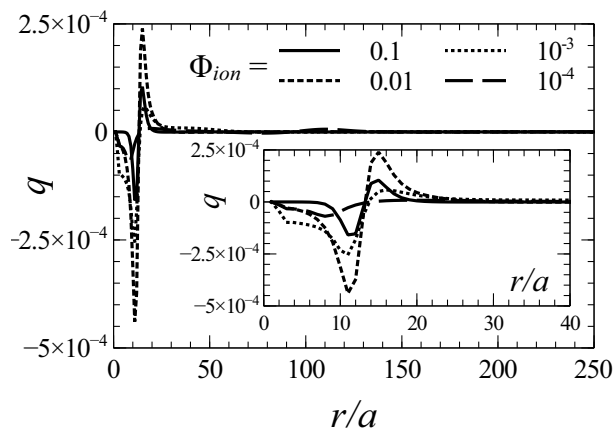}
\caption{\label{fig:SCF_qtot}Radial distribution of local charge in AB-core-shell star with $p=20$, $N_{A}=1000$, $N_{B}=200$, $\chi_{A}=0$, $\chi_{B}=1$, $\alpha=0.5$ and different
values of $\phi_{ion}$.}
\end{figure}

Finally, to test the validity of the corona electroneutrality assumption 6, local charge density distribution $q(r)=\varphi_{Na^+}(r) + \alpha_{b}\varphi_{A}(r) - \varphi_{Cl^-}(r)$ was calculated. Figure \ref{fig:SCF_qtot} shows that in the corona, the local charge density is perfectly zero, the distortions of local electroneutrality are observed at the A/B interface only.

\section{Conclusions and outlook}
In the present paper, conformational transitions in an amphiphilic
(AB)$_{p}$ block copolymer star with hydrophobic central blocks and polyelectrolyte
terminal blocks were studied by means of analytical mean-field theory combined with the SF-SCF numerical modeling. 
The star is in aqueous media and forms under high ionic strength conditions
a unimolecular micelle with collapsed B-core and extended
A corona. Because of repulsive electrostatic interactions in the polyelectrolyte corona
the core is subjected to the radial extensional force. The strength
of the interactions in the corona is governed by the degree of ionization
of A blocks and by the ionic strength of the solution.
From the experimental viewpoint, the salt concentration is the most common
control parameter.
%from the practical viewpoint, it is most convenient to chose the latter
%as a control parameter.
We have shown that a decrease in the ionic strength
of the solution may provoke the intra-molecular conformational transition related to the unfolding of collapsed inner hydrophobic blocks.
More specifically, the unimolecular (AB)$_{p}$ micelle undergoes two consecutive
conformational transitions: The first one is continuous and is associated to the onset
of the microphase segregation in the B-core. 
%requires the A-corona
%pulling force to be greater than a certain threshold), 
The second
one is the complete unfolding of the core which is accompanied
by the jumpwise increases of the core size and the overall star size
and the corresponding jumpwise decrease in the corona thickness. 
Remarkably, the second transition does not occur in the stars with sufficiently
large number of arms.
We
have constructed the diagrams of states of (AB)$_{p}$ star in $\Phi_{ion}-\alpha_{b}$
and $\Phi_{ion}-p$ coordinates. The diagrams contain the region corresponding to 
unimolecular micelles with collapsed core, the region corresponding to the partial unfolded 
core-forming blocks and, finally, the region corresponding to 
the conformation with completely unfolded core and extended B blocks.
The developed analytical theory is based on two essential approximations:
The first one is the approximation of local electroneutrality of the corona.
Both analytical and numerical Poisson-Boltzmann analysis convincingly prove \cite{Borisov:2011} 
that for reasonable degree of ionization ($\alpha_b\geq 0.1$) the approximation is very accurate for
the stars with the number of arms $p\geq 10$.
%(see for discussion \cite{Borisov:2011}) 
%%%%%%%%
%\bibitem{Borisov_et_al stars} Borisov OV, Zhulina EB, Leermakers FAM, M\"uller
%AHE, Ballauff M Adv Polym Sci, 2011, DOI: 10.1007/12_2010_104
%%%%%%%%
The second one is the equal arm stretching approximation. The latter is well-justified
for the unimolecular micelles with collapsed B-core as well as for the conformations with completely
unfolded core~ \cite{Borisov:2011}. The situation, however, is more delicate
in the regime with phase segregated (partially unfolded) core. In particular, one could expect that there is
an analogy to the collapse transition in hydrophobic polyelectrolyte stars and brushes: Theories based on
the equal stretching approximation \cite{Borisov:1991, Borisov:1992, Ross:1992}
%%%%%%%%%%%%%%%%%%%%
%\bibitem{Borisov_Birshtein_Zhulina_JdP1991} Borisov OV, Birshtein TM,Zhulina EB (1991) J Phys II (France)1: 521
%
%\bibitem{Borisov_Birshtein_Zhulina_PCPS1992} Borisov OV, Birshtein TM,Zhulina EB (1992) Progress in Colloid and Polymer Science 90: 177
%
%\bibitem{Pincus_Ross_Mm1992} Ross R, Pincus P (1992) Macromolecules 25: 2177.
%%%%%%%%%%%%%%%%%%%%%
predicted a jump-wise "unfolding" transition. 
On the other hand, more accurate approaches~\cite{Misra:1994, Pryamitsyn:1996}
%%%%%%%%%%%%%%
%\bibitem{Misra_Mattice_Napper1994} Misra S, Mattice WL, Napper DH (1994) Macromolecules 27: 7090
%\bibitem{Pryamitsyn}Pryamitsyn V.A., Leermakers F.A.M., Fleer G.J., Zhulina E.B., Macromolecules, 1996, 29, 8260
%\bibitem{Polotsky et al} Polotsky et al, in preparation
%%%%%%%%%%%%%%
pointed to the co-existence of two populations of chains (collapsed and stretched) and smooth variation
of the average dimensions in the transition range due to progressive re-partitioning of the chains
between the collapsed and the extended states. 
Therefore, we have complemented our analytical theory of the unfolding transition in the (AB)$_{p}$ block copolymer stars 
by numerical calculations made
with the aid of assumption-free SF-SCF method. The results of the
SCF modeling have confirmed the predicted scenario of the core unfolding
in the amphiphilic (AB)$_{p}$ star and shown that with an increase
in electrostatic repulsions in the corona, the core unfolding occurs via microphase
segregated states. Importantly, the SCF calculations have proven that at any ionic strength
(i.e., both in the microphase-segregated and and in the unfolded conformations) the end segments
of the A-blocks and the A-B junction points exhibit unimodal radial distributions, each with a single maximum.
This observation proves that the analytical theory based on the equal stretching approximation
gives a qualitatively correct (artifact-free) description of the conformations and intra-molecular 
unfolding transition in amphiphilic (AB)$_{p}$ star. 

In the present paper we considered (AB)$_{p}$ star copolymers with a quenched fraction of permanently charged monomer units
in the A-block. Correspondingly, the ionic strength of the solution was chosen as a control parameter, variation in which triggers
the intra-molecular conformational transitions. In the case of the copolymers comprising weakly dissociating (pH-sensitive) polyelectrolyte
blocks, the degree of ionization and, as a result, the conformation of the copolymer can be affected by variation in pH as well.
Moreover, in the case of branched polyelectrolytes the degree of ionization of branches is not controlled solely by the value of pH in the buffer,
but depends also on the ionic strength of the solution and on the polymer conformation. This coupling causes highly non-trivial responsive behavior of the
pH-sensitive branched (co)polymers.
In particular, continuous variation in the ionic strength 
at pH$\approx$pK may
cause re-entrant folding-unfolding transition. These effects will be addressed in the forthcoming publication.

Above we have considered very dilute solutions of (AB)$_{p}$ star copolymers
which form spherical unimolecular micelles.
At higher concentrations, the aggregation
of (AB)$_{p}$ unimolecular micelles with small number of arms may lead to formation of 
intermolecular assemblies. Depending on the lengths of the hydrophobic and polyelectrolyte blocks
this aggregation may give rise not only to spherical, but also to cylindrical micelles, lamellar or bicontinuous
structure~\cite{Strandman:2007}. 
% {\bf Here we may give a reference to the experiment by Satu Strandman: 
% Strandman, S.; Zarembo, A.; Darinskii, A.; L?flund, B.; Butcher, S.J.; Tenhu, H., 
% Polymer 2007, 48, 7008}
For amphiphilic ionic/hydrophobic diblock copolymers, the morphology of the inter-molecular aggregates is controlled by the length of the
A and B blocks, the morphological transitions can be triggered by variation in the ionic strength or in pH
%(see~\cite{Borisov:2011} for the review).
%%%%%%%%
%\bibitem{Borisov_et_al micelles} Borisov OV, Zhulina EB, Leermakers FAM, M\"uller
%AHE, Adv Polym Sci, 2011, DOI: 10.1007/12_2010_114
%%%%%%%%
The connectivity of the amphiphilic block copolymers into stars introduces additional topological constraints 
which may influence the morphologies of the inter-molecular assemblies of (AB)$_{p}$ stars as compared to those formed by homologous 
diblock copolymers ("single arms"). 
We remark, however, that because of strong steric repulsions between polyelectrolyte coronae of the unimolecular micelles,
their aggregation may be significantly hindered kinetically.
Analysis of the inter-molecular assembly of the (AB)$_{p}$ star copolymers is, however, beyond the scope of the present work.
% {\bf I would remove the following paragraph (see the TeX file), but check 
% if nothing important is lost}
%To describe radial extensional deformation of the collapsed core, the approach
%developed earlier in \cite{Polotsky:2010} to study mechanical unfolding
%of a polymer globule formed by a single flexible linear macromolecule
%was used. It was assumed that the core blocks are equally extended
%so that their ends (A-B junction points) are placed at the distance
%$R$ from the center of the star. When $R$ exceeds the radius of
%unperturbed star, microphase segregation occurs: part of each block
%leaves the core and form stretched {}``leg'' while the rest remains
%in the collapsed core (globule). A general picture of star extension
%is similar to that of a globule formed by a linear macromolecule (formally
%this correspond to the star with $p=2$): with an increase in $R$
%the tail length grows, the core size decreases and the reaction force
%slightly decreases (quasi-plateau). When the extension reaches a certain
%threshold, a complete jumpwise unfolding (unraveling) of the core
%occurs. Increasing the number of arms $p$ the character of the force-deformation
%dependence in the quasi-plateau regime changes its trend and becomes
%a weakly increasing one. At the same time the jump in the reaction
%force decreases, and the core unfolding becomes a continuous transition.
%The PE corona free energy was calculated in the framework of the local
%(or Daoud-Cotton) model of polymer brush.
%%%%%%%%%%%%%%%%%%%%%%%%%%
\section*{Acknowlegdement}
\label{Acknowlegdement}
%%%%%%%%%%%%%%%%%%%%%%%%%%
This work has been performed as a part of
the collaborative research project SONS-AMPHI within the European
Science Foundation EUROCORES Program, and has been
partially supported by funds from the EC Sixth Framework Program
through the Marie Curie Research and Training Network POLYAMPHI.
Support by the Dutch National Science Foundation
(NOW) and the Russian Foundation for Basic Research (RFBR)
through Joint Project 047.017.026/06.04.89402 and Project 08-03-00336a
is gratefully acknowledged.
The authors are thankful to Professor F.A.M.Leermakers for introduction to the
SF-SCF approach and sharing software package {\em SFBox}.
Financial support by the Russian Foundation for Basic Research (RFBR) through Project 11-03-00969-a, 
by the Department of Chemistry and Material Science of the Russian Academy of Sciences,
by the funds from the EC Sixth Framework Program through the Marie Curie Research and Training Network
POLYAMPHI
and by the 
Scientific and Technological Cooperation
Program Switzerland - Russia, project \textquotedblleft Experimental studies
and theoretical modeling of amphiphilic di/triblock and dendritic functional
polymers at surfaces: influence of interfacial architecture on biological
response\textquotedblright , Grant Agreement No.128308.
are gratefully acknowledged. 

\bibliographystyle{unsrt}
\bibliography{Core-Shell-Star}
\end{document}